\documentclass[preprint,12pt]{elsarticle}
\usepackage{amssymb}
\usepackage{amsthm}
\usepackage{framed} % To put a box to the nomenclature
\usepackage{amsmath,color}
\usepackage{mathrsfs}
\usepackage{graphicx}
\usepackage{epstopdf}
\usepackage{float}
\usepackage{caption}
\usepackage{subcaption}
\usepackage{bm}
\usepackage{bbm}
\usepackage{mathrsfs}
\usepackage{cleveref}
\usepackage{soul}
\usepackage{accents}
\usepackage{color,soul} %to highlight text
\usepackage{color} %To highlight references in the text
\usepackage{nomencl} %to add the nomenclature
\makenomenclature %to add the nomenclature
\biboptions{sort&compress}
\soulregister\citep7 % To allow citep to be inside of highlighted text
\soulregister\citet7 % Idem with citet
\soulregister\citealp7 % Idem with citealp
\newsavebox{\measurebox} %To create a 1 column figure with a 2 column figure on its side
\usepackage{titlesec} %To use paragraph as subsubsubsection
\usepackage{tabu} % To increase the vertical space between cells in tables
\usepackage{longtable}
\usepackage{tikz}

\journal{}

\makeatletter
\def\@author#1{\g@addto@macro\elsauthors{\normalsize%
    \def\baselinestretch{1}%
    \upshape\authorsep#1\unskip\textsuperscript{%
      \ifx\@fnmark\@empty\else\unskip\sep\@fnmark\let\sep=,\fi
      \ifx\@corref\@empty\else\unskip\sep\@corref\let\sep=,\fi
      }%
    \def\authorsep{\unskip,\space}%
    \global\let\@fnmark\@empty
    \global\let\@corref\@empty  %% Added
    \global\let\sep\@empty}%
    \@eadauthor={#1}
}
\makeatother

\titleformat{\paragraph}
{\normalfont\normalsize\itshape}{\theparagraph}{1em}{}
\titlespacing*{\paragraph}
{0pt}{3.25ex plus 1ex minus .2ex}{1.5ex plus .2ex}

\begin{document}

\begin{frontmatter}

%% Title, authors and addresses

%% use the tnoteref command within \title for footnotes;
%% use the tnotetext command for theassociated footnote;
%% use the fnref command within \author or \address for footnotes;
%% use the fntext command for theassociated footnote;
%% use the corref command within \author for corresponding author footnotes;
%% use the cortext command for theassociated footnote;
%% use the ead command for the email address,
%% and the form \ead[url] for the home page:
%% \title{Title\tnoteref{label1}}
%% \tnotetext[label1]{}
%% \author{Name\corref{cor1}\fnref{label2}}
%% \ead{email address}
%% \ead[url]{home page}
%% \fntext[label2]{}
%% \cortext[cor1]{}
%% \address{Address\fnref{label3}}
%% \fntext[label3]{}

\title{Size effects in elastic-plastic functionally graded materials}
%\title{Non-local plasticity modeling of metallic functionally graded materials}

%% use optional labels to link authors explicitly to addresses:
%% \author[label1,label2]{}
%% \address[label1]{}
%% \address[label2]{}

\author{Tittu V. Mathew\fnref{IITM}}

\author{Sundararajan Natarajan\fnref{IITM}}

\author{Emilio Mart\'{\i}nez-Pa\~neda\corref{cor1}\fnref{Cam}}
\ead{mail@empaneda.com}

\address[IITM]{Department of Mechanical Engineering, Indian Institute of Technology - Madras, Chennai - 600036, India}

\address[Cam]{Department of Engineering, Cambridge University, CB2 1PZ Cambridge, UK}

\cortext[cor1]{Corresponding author.}

\begin{abstract}
We develop a strain gradient plasticity formulation for composite materials with spatially varying volume fractions to characterize size effects in functionally graded materials (FGMs). The model is grounded on the mechanism-based strain gradient plasticity theory and effective properties are determined by means of a linear homogenization scheme. Several paradigmatic boundary value problems are numerically investigated to gain insight into the strengthening effects associated with plastic strain gradients and geometrically necessary dislocations (GNDs). The analysis of bending in micro-size functionally graded foils shows a notably stiffer response with diminishing thickness. Micro-hardness measurements from indentation reveal a significant increase with decreasing indenter size. And large dislocation densities in the vicinity of the crack substantially elevate stresses in cracked FGM components. We comprehensively assess the influence of the length scale parameter and material gradation profile to accurately characterize the micro-scale response and identify regimes of GNDs relevance in FGMs.
\end{abstract}

\begin{keyword}

Strain gradient plasticity \sep Functionally graded materials \sep Micro-scale plasticity \sep Finite element analysis \sep Fracture
%% keywords here, in the form: keyword \sep keyword

%% PACS codes here, in the form: \PACS code \sep code

%% MSC codes here, in the form: \MSC code \sep code
%% or \MSC[2008] code \sep code (2000 is the default)

\end{keyword}

\end{frontmatter}

%% \linenumbers

%% main text

\section{Introduction}
\label{Sec:Introduction}
Functionally graded materials (FGMs) are multifunctional composites with spatially varying volume fractions of constituent materials. The resulting graded macroproperties enable designers to tailor the microstructure to specific operating conditions, while minimizing problems associated with discrete material interfaces. FGMs are widely used in the biomedical, aerospace, and automotive sectors, and have become particularly popular in microelectromechanical systems (MEMS) \cite{Fu2004}. Conventional continuum theories are limited when dealing with micro and nano-sized functionally graded components and, as a consequence, non-local approaches have been extensively used to accurately characterize their mechanical behavior \cite{Natarajan2012,Natarajan2014,Lou2016,Simsek2017,Sahmani2018,Momeni2018}. However, despite the fact that metal-metal FGMs dominate MEMS and microelectronic applications, these studies limit their analyses to the elastic response. Elasto-plastic investigations of deformation and fracture in FGMs have soared in recent years (see, e.g., \cite{Tsiatas2017,Amirpour2017}) but they are still confined to conventional length-independent plasticity. Numerous micron scale experiments have shown that metallic materials display strong size effects when deformed non-uniformly into the plastic range. Particularly representative examples are indentation \cite{Nix1998}, torsion \cite{Fleck1994}, and bending \cite{Stolken1998}. The \emph{smaller is stronger} response observed is attributed to the work hardening contribution of geometrically necessary dislocations (GNDs) that arise in the presence of plastic strain gradients to ensure geometric compatibility. A notable effort has been devoted to extend classic plasticity to the small scales by the development of strain gradient plasticity (SGP) models, which define the plastic work as a function of both strains and strain gradients \cite{Gao1999,Fleck2001,Huang2004,Gurtin2005}. By introducing an intrinsic length scale in the constitutive equations, SGP theories are able to quantitatively capture the size-dependent response of metals. Mao et al. \cite{Mao2013} recently showed that size effects can significantly impact the bending response of graded materials but they assumed a constant length scale parameter throughout the beam. We comprehensively investigate the influence of plastic strain gradients in deformation and fracture of FGMs. The mechanism-based SGP theory is extended to graded materials by defining elastic and plastic properties as a function of the material volume fraction through an appropriate homogenization scheme. The analysis of bending, indentation and stationary cracks reveals a profound GND-effect and a strong sensitivity of the results to the profile of the length scale parameter.

\section{A size-dependent plasticity formulation for FGMs}
\label{Sec:NumModel}

We develop a mechanism-based model for graded materials that incorporates the role of geometrically necessary dislocations (GNDs) through a Taylor-based flow stress \cite{Gao1999,Huang2004}. We consider a metal-metal FGM specimen that gradually changes from 100\% volume fraction of titanium to 100\% volume fraction of aluminum. Assuming an FGM beam with thickness $h$ and material gradation along a $y$-axis centered at the mid-plane, the volume fraction of material 1, $V_1$, reads,
\begin{equation}
V_1 = \left(\frac{1}{2} + \frac{y}{h} \right)^k
\end{equation}

\noindent where $k$ is the material gradient index or volume fraction exponent.

\subsection{Homogenization scheme}
\label{Sec:FGM}

A Mori-Tanaka homogenization scheme is employed to obtain the local effective elastic properties as a function of the volume fraction. Thus, the effective bulk modulus $K_e$ and shear modulus $\mu_e$ can be obtained as,
\begin{equation}
\frac{K_e-K_1}{K_2 - K_1} = \frac{V_2}{1+3 V_1  \left(K_2 - K_1 \right) / \left( 3 K_1 + 4 \mu_1 \right)}
\end{equation}
\begin{equation}
\frac{\mu_e-\mu_1}{\mu_2-\mu_1} = \frac{V_2}{1+V_1 \left(\mu_2 - \mu_1 \right)/\left(\mu_1 + \mu_1 \left( 9 K_1 + 8 \mu_1 \right) / \left( 6 \left( K_1 + 2 \mu_1 \right) \right) \right)}
\end{equation}

\noindent and one can readily compute the effective Young's modulus $E_e$ and Poisson's ratio $\nu_e$ from $K_e$ and $\mu_e$ using the standard relations.\\

Furthermore, we compute the local effective yield stress ${\sigma_Y}_e$ and the effective strain hardening exponent $N_e$ using the rule of mixtures. E.g., the effective yield stress is given by,
\begin{equation}
{\sigma_Y}_e = {\sigma_Y}_1 V_1 + {\sigma_Y}_2 V_2
\end{equation}

\noindent where ${\sigma_Y}_1$ and ${\sigma_Y}_2$ respectively denote the yield strength of material 1 and material 2. 

\subsection{Taylor-based dislocation hardening}
\label{Sec:Theory}

The size-dependent response of metals is captured by means of a mechanism-based approach that builds upon Taylor dislocation model (see \cite{Gao1999,Huang2004}). Thus, the shear flow stress $\tau$ is formulated in terms of the total dislocation density $\rho$ as,
\begin{equation}\label{Eq:tau}
\tau = \alpha \mu_e b \sqrt{\rho}
\end{equation}

\noindent Here, $b$ is the magnitude of the Burgers vector and $\alpha$ is an empirical coefficient that is generally taken to be 0.5. The mechanism-based strain gradient (MSG) plasticity model provides an implicit multiscale framework where the microscale concepts of statistically stored dislocations (SSDs) and geometrically necessary dislocations (GNDs) are respectively linked to the mesoscale concepts of plastic strains and plastic strain gradients. Accordingly, the dislocation density is composed of the SSDs density, $\rho_S$, and the GNDs density, $\rho_G$, 
\begin{equation}
\rho = \rho_S + \rho_G
\end{equation}

The GND density $\rho_G$ is related to the effective plastic strain gradient by,
\begin{equation}\label{Eq:rhoG}
\rho_G = \bar{r} \frac{\eta^p}{b}
\end{equation}

\noindent where $\bar{r}$ is the Nye-factor which is taken to be 1.9 for face-centered-cubic (fcc) polycrystals. The effective plastic strain gradient is defined as,
\begin{equation}
\eta^p = \sqrt{\frac{1}{4} \eta^p_{ijk} \eta^p_{ijk}}
\end{equation}

\noindent where the third-order tensor $\eta^p_{ijk}$ is given by,
\begin{equation}
\eta^p_{ijk} = \varepsilon^p_{ik,j} + \varepsilon^p_{jk,i} - \varepsilon^p_{ij,k}
\end{equation}

The tensile flow stress $\sigma_{flow}$ is related to the shear flow stress $\tau$ by,
\begin{equation}
\sigma_{flow}= M \tau
\end{equation}

\noindent where $M$ is the Taylor factor, taken to be 3.06 for fcc metals. Rearranging Eqs. (\ref{Eq:tau})-(\ref{Eq:rhoG}) yields,
\begin{equation}\label{Eq:Sflow}
\sigma_{flow}=M \alpha \mu_e b \sqrt{\rho_S + \bar{r} \frac{\eta^p}{b}}
\end{equation}

The SSD density $\rho_S$ can be determined from (\ref{Eq:Sflow}) knowing the relation in uniaxial tension between the flow stress and the material stress-strain curve,
\begin{equation}
\rho_S = \left( \frac{\sigma_{ref} f \left( \varepsilon^p \right)}{M \alpha \mu_e b} \right)^2
\end{equation}

Here, $\sigma_{ref}$ is a reference stress and $f$ is a non-dimensional function of the plastic strain $\varepsilon^p$ determined from the uniaxial stress-strain curve. Substituting back into (\ref{Eq:Sflow}), $\sigma_{flow}$ yields
\begin{equation}
\sigma_{flow} = \sigma_{ref} \sqrt{f^2 \left( \varepsilon^p \right) + \ell_e \eta^p}
\end{equation}

\noindent where $\ell_e$ is the (effective) intrinsic material length, which depends on parameters of elasticity ($\mu_e$), plasticity ($\sigma_{ref}$), and atomic spacing ($b$),
\begin{equation}\label{Eq:length}
\ell_e = M^2 \bar{r} b \alpha^2 \left( \frac{\mu_e}{\sigma_{ref}}\right)^2 = 18 \alpha^2 b \left( \frac{\mu_e}{\sigma_{ref}}\right)^2 
\end{equation}

Here, we assume the following isotropic hardening law,
\begin{equation}
\sigma = {\sigma_Y}_e \left( 1 + \frac{E_e \varepsilon^p}{{\sigma_Y}_e} \right)^{N_e}
\end{equation}

\noindent where we define the reference stress as $\sigma_{ref}={\sigma_Y}_e \left( E_e/ {\sigma_Y}_e \right)^{N_e}$ and $f \left( \varepsilon^p\right)= \left( \varepsilon^p + {\sigma_Y}_e/E_e \right)^{N_e}$.

\subsection{Finite element implementation}
\label{Sec:ABAQUS}

We follow the work by Huang et al. \cite{Huang2004} and adopt a lower order implementation due to its robustness in finite strain problems (see \cite{Hwang2003,IJSS2015}). As discussed in \cite{Qu2004,TAFM2017} for homogeneous materials, Taylor's dislocation model relates the flow stress to both strains and strain gradients,
\begin{equation}
\dot{\sigma} = \frac{\partial \sigma}{\partial \varepsilon^p} \dot{\varepsilon}^p + \frac{\partial \sigma}{\partial \eta^p} \dot{\eta}^p
\end{equation}

\noindent requiring higher order stresses to render a self contained constitutive model. This can be overcome in a conventional setting by using a viscoplastic formulation that particularizes to the rate independent limit. Thus, we define the plastic strain rate as a function of the effective stress $\sigma_e$,
\begin{equation}
\dot{\varepsilon}^p = \dot{\varepsilon} \left( \frac{\sigma_e}{ \sigma_{ref} \sqrt{f^2 \left( \varepsilon^p \right) + \ell_e \eta^p}} \right)^m
\end{equation}

\noindent and substitute the reference strain by an effective strain rate $\dot{\varepsilon}$. Values of $m$ larger than 20 have proven to capture the rate independent response \cite{Huang2004}. The field equations are therefore the same as in conventional plasticity theories and $\eta^p$ comes into play through the incremental plastic modulus. The constitutive equation reads,
\begin{equation}
\dot{\sigma}_{ij}= K_e \dot{\varepsilon}_{kk} \delta_{ij} + 2 \mu_e \left( \dot{\varepsilon}_{ij}' - \frac{3 \dot{\varepsilon}}{2 \sigma_e} \left(\frac{\sigma_e}{\sigma_{flow}} \right)^m \dot{\sigma}_{ij} \right)
\end{equation}

\noindent with $\delta_{ij}$ being the Kronecker delta and $\sigma_{ij}$ the Cauchy stress tensor. Both lower and higher order versions of MSG plasticity predict identical results in their physical regime of validity \cite{Qu2004}.\\

The size-dependent plasticity formulation for FGMs outlined is implemented in the well-known finite element package ABAQUS by means of a UMAT subroutine. The plastic strain gradient is obtained by numerical differentiation within the element: the plastic strain increment is interpolated through its values at the Gauss points in the isoparametric space and afterwards the increment in the plastic strain gradient is calculated by differentiation of the shape functions. In the present finite strain analysis, rigid body rotations for the strains and stresses are carried out by means of the Hughes and Winget’s algorithm \cite{Hughes1980a} and the strain gradient is obtained from the deformed configuration. Post-processing of the results is carried out by means of Abaqus2Matlab \cite{AES2017}.

\section{Results}
\label{Sec:Results}

We comprehensively investigate the behavior of an aluminum-titanium FGM under bending (Section \ref{Sec:Microbending}), indentation (Section \ref{Sec:Indentation}) and fracture (Section \ref{Sec:Fracture}). Applications of Al-Ti FGMs include jet engine components and dynamic high-pressure technology \cite{Mortensen1995,Xiong2000}. The volume fraction of material changes gradually along the $y$ axis, with material properties at the edges given by Table \ref{Tab:MatProperties}.

\begin{table}[H]
\centering
\caption{Material properties of the metal-metal FGM under consideration.}
\label{Tab:MatProperties}
   {\tabulinesep=1.2mm
   \begin{tabu} {cccccc}
       \hline
 Material & E (GPa) & $\nu$ & $\sigma_Y$ (MPa) & N & $\ell$ ($\mu$m) \\ \hline
 Aluminum   & 69 & 0.334 & 276 & 0.042 & 6.23 \\
 Titanium & 113.8 & 0.342 & 880 & 0.21 & 0.34 \\\hline
   \end{tabu}}
\end{table}

\begin{figure}[H] 
    \centering
    \includegraphics[scale=1]{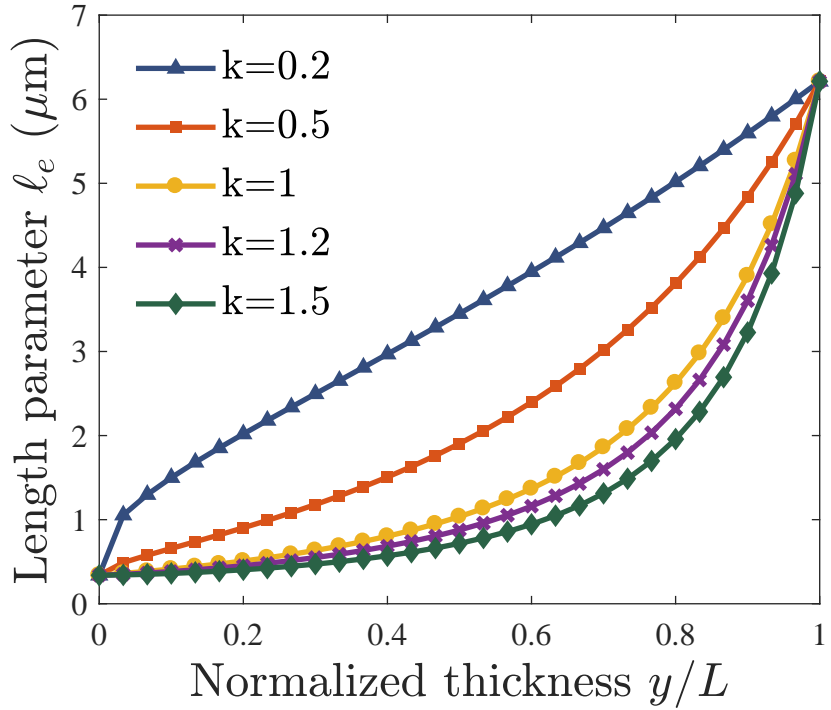}
    \caption{Variation of the effective material length scale $\ell_e$ along the material gradient direction.}
    \label{fig:LengthScaleVar}
\end{figure}

The effect of geometrically necessary dislocations in the mechanical response is accounted for by means of a mechanism-based strain gradient plasticity formulation for graded materials. In addition, we take into consideration the variation of the length scale parameter through the thickness. Thus, the effective material length is given by (\ref{Eq:length}) and its variation along the specimen characteristic dimension is illustrated for different values of the material gradient index $k$ in Fig. \ref{fig:LengthScaleVar}. As shown in the figure, the length scale parameter changes gradually from the $\ell=0.34$ $\mu$m value for titanium to the $\ell=6.23$ $\mu$m value for aluminum; increasing $k$ entails smaller average values of $\ell$. We investigate the influence of the spatial variation of the length scale parameter by considering four scenarios: (i) $\ell$ equals the Al value ($\ell_{MAX}$), (ii) $\ell$ equals the Ti value ($\ell_{MIN}$), (iii) $\ell=0$ (conventional plasticity), and (iv) $\ell$ changes gradually according to (\ref{Eq:length}).

\subsection{Micro-bending of thin FGM films}
\label{Sec:Microbending}

\begin{figure}[H]
\centering
\begin{tikzpicture}[thick]
\shade[top color=teal,bottom color=white, draw=black] (0,0) rectangle (10,3);
\draw[very thick,->] (0,1.5) -- (4,1.5) node[anchor=north west] {x axis};
\draw[very thick,->] (0,1.5) -- (0,4) node[anchor=south east] {y axis};
\node at (2,0.5) {Titanium};
\node at (2,2.5) {Aluminum};
\foreach \y in {0,0.8,2.2,3}
	\draw[black,thick] (-0.2,\y) circle (0.2);
\draw (0,1.5) -- (-0.4,1.8) --(-0.4,1.2) -- cycle;
\foreach \y in {1.75,2,2.25,2.5,2.75,3}
	\draw[black,very thick,->] (10,\y) -- (8.5+\y,\y); 
\foreach \y in {0,0.25,0.5,0.75,1,1.25}
	\draw[black,very thick,->] (10,\y) -- (8.5+\y,\y); 
\draw[black,very thick,<->] (0,3.5) -- (10,3.5); 
\draw[black,very thick,-] (10,3.25) -- (10,3.75);
\draw[black,very thick,<->] (-1,0) -- (-1,3); 
\draw[black,very thick,-] (-1.25,0) -- (-0.75,0);
\draw[black,very thick,-] (-1.25,3) -- (-0.75,3);
\node at (5,4) {W/2};
\node at (-1.5,1.5) {H};
\end{tikzpicture}
\caption{Schematic representation of the numerical micro-bending experiments on thin metal-metal FGM films.} \label{fig:MicroBend}
\end{figure}
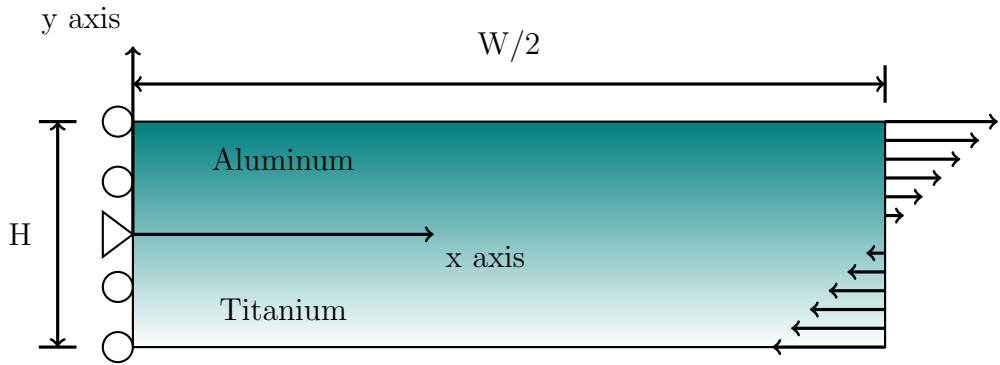

Bending of micron-sized foils is a paradigmatic benchmark of small scale plasticity since the experiments of St\"{o}lken and Evans \cite{Stolken1998} (see, e.g., \cite{Idiart2009,IJSS2016} and references therein). We model a functionally graded foil of thickness $H$ and length $W$ subjected to bending. Unless otherwise stated, a foil thickness of $H=0.01$ mm is considered. As shown in Fig. \ref{fig:MicroBend}, we take advantage of symmetry and model only half of the beam. The longitudinal displacement component is prescribed at the beam end ($x \pm W/2$) as,
\begin{equation}
u = \kappa \, x \, y 
\end{equation}

\noindent where $\kappa$ is the applied curvature. The beam is uniformly meshed with a total of 12000 plane strain quadrilateral quadratic elements with reduced integration. After a sensitivity study, 40 elements are employed along the beam thickness. Material properties change gradually along the $y$-axis from aluminum (top) to titanium (bottom).

\begin{figure}[H]
\makebox[\linewidth][c]{%
        \begin{subfigure}[b]{0.6\textwidth}
                \centering
                \includegraphics[scale=0.67]{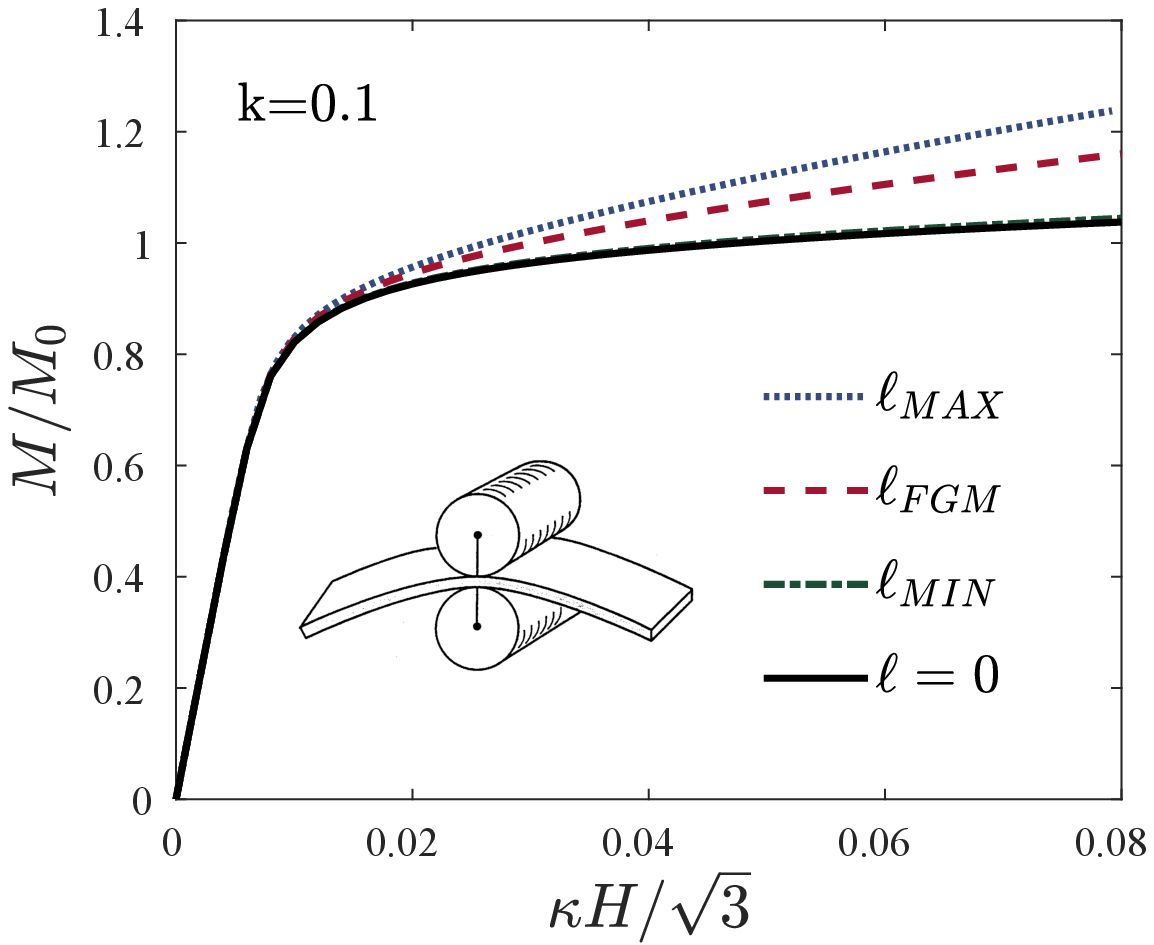}
                \caption{}
                \label{fig:MvskappaA}
        \end{subfigure}
        \begin{subfigure}[b]{0.6\textwidth}
                \raggedleft
                \includegraphics[scale=0.67]{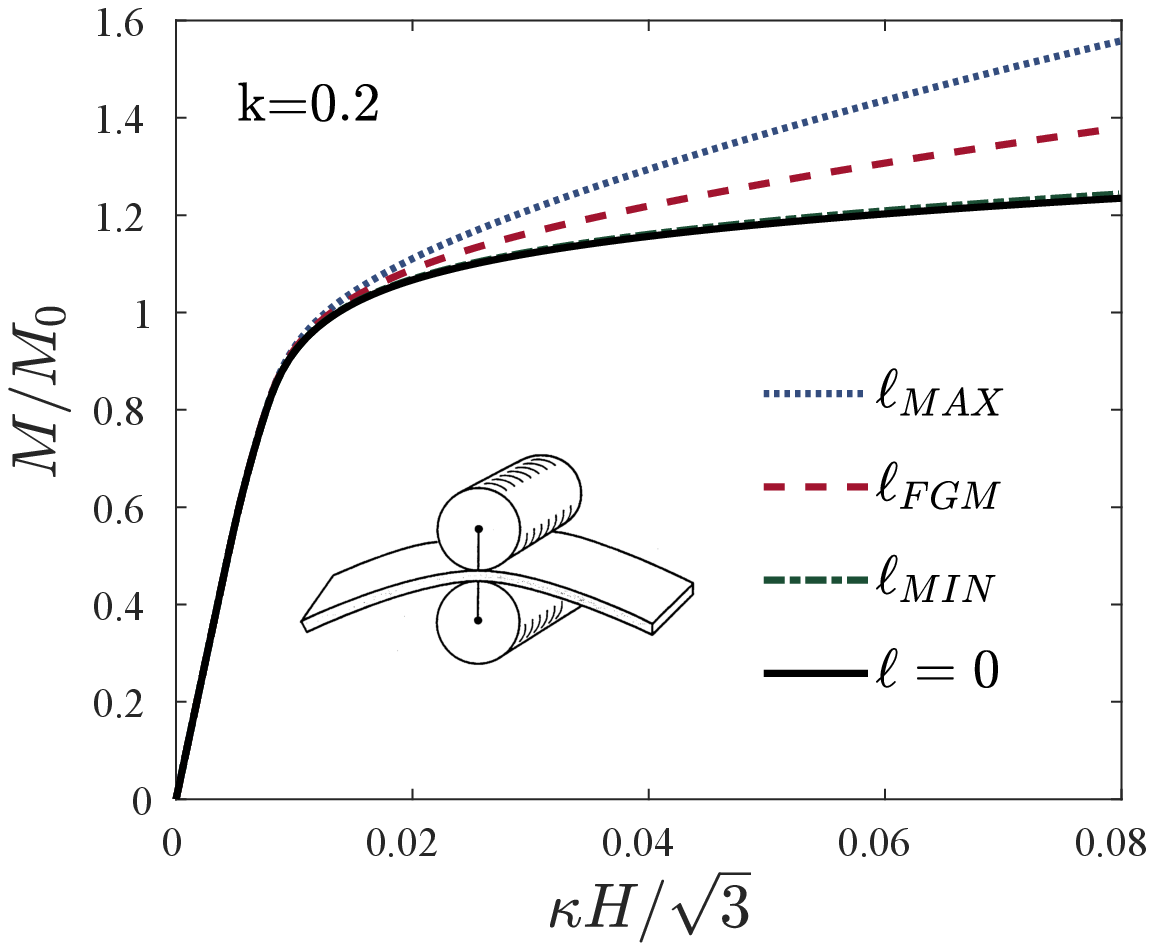}
                \caption{}
                \label{fig:MvskappaB}
        \end{subfigure}}

\makebox[\linewidth][c]{%       
        \begin{subfigure}[b]{0.6\textwidth}
                \centering
                \includegraphics[scale=0.67]{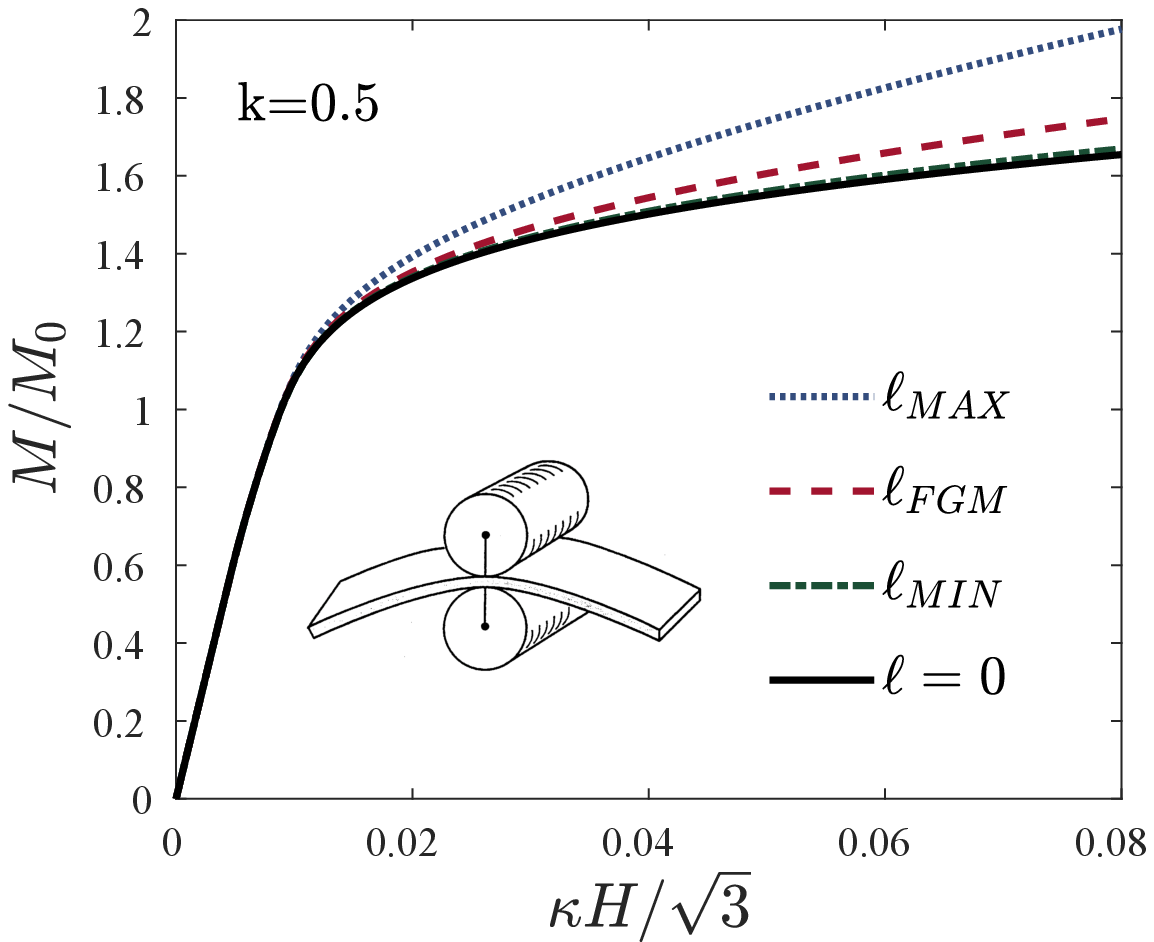}
                \caption{}
                \label{fig:MvskappaC}
        \end{subfigure}
        \begin{subfigure}[b]{0.6\textwidth}
                \raggedleft
                \includegraphics[scale=0.67]{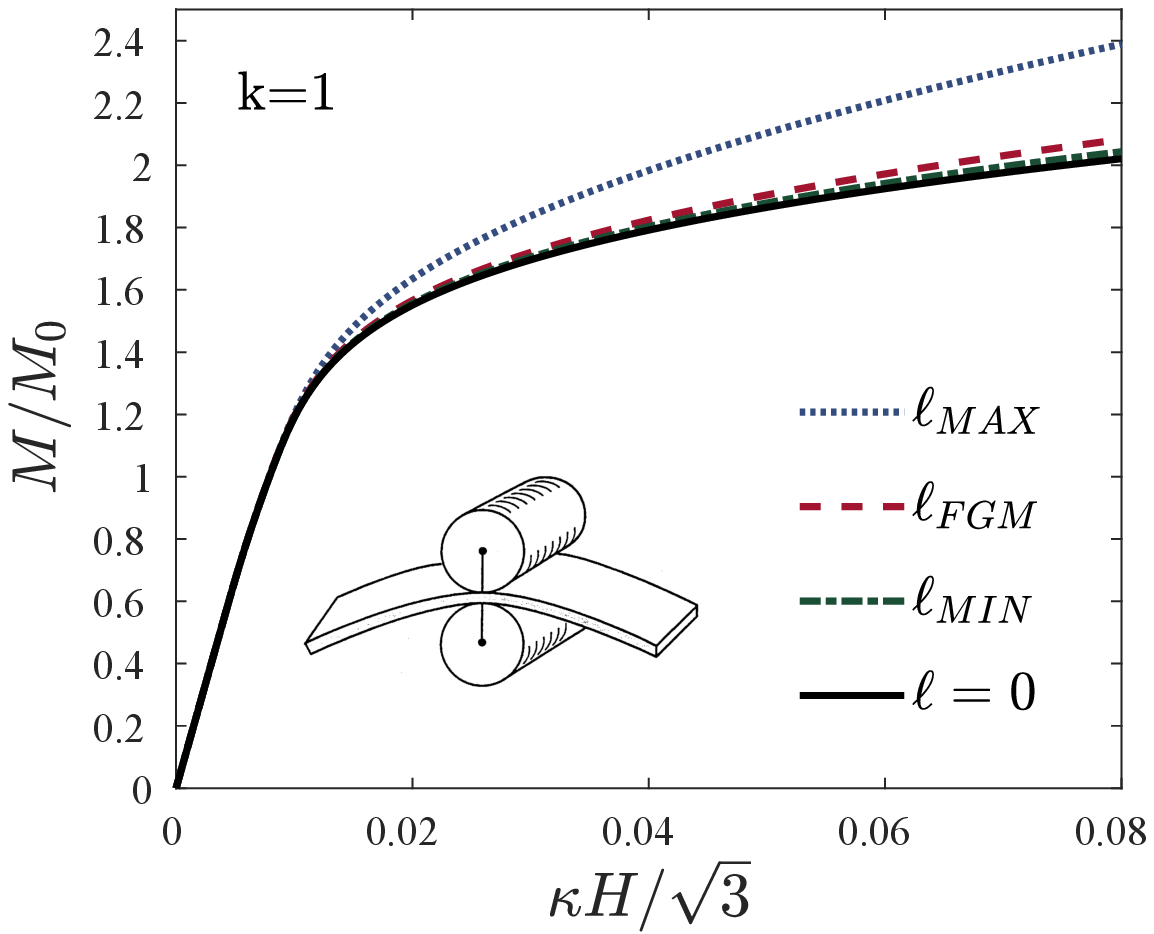}
                \caption{}
                \label{fig:MvskappaD}
        \end{subfigure}
        }       
        \caption{Normalized moment versus curvature for conventional ($\ell=0$) and strain gradient plasticity theories. A functionally graded length parameter ($\ell_{FGM}$) has been considered, along with the limit cases of the stiffer ($\ell_{MAX}$) and the softer ($\ell_{MIN}$) compounds. Material gradient indexes: (a) $k=0.1$, (b) $k=0.2$, (c) $k=0.5$, and (d) $k=1$.}\label{fig:Mvskappa}
\end{figure}

We compute the bending moment and normalize by,
\begin{equation}
M_0 = \frac{\bar{\sigma}_Y H^2}{6 \sqrt{1 - \bar{\nu} + \bar{\nu}^2}}
\end{equation}

\noindent where $\bar{\sigma}_Y$ and $\bar{\nu}$ correspond to the average yield stress (578 MPa) and Poisson's ratio (0.338), respectively. Unlike the cases of indentation (Section \ref{Sec:Indentation}) and fracture (Section \ref{Sec:Fracture}), the use of the infinitesimal deformation theory is appropriate for bending, given the strain levels attained. Results are shown in Fig. \ref{fig:Mvskappa} as a function of the normalized applied curvature for different values of the material gradient index. In terms of the length scale parameter, we assess the role of an accurate characterization of $\ell$ in the moment versus curvature response by considering the four scenarios outlined above.

Results reveal a number of trends. First, the expected \emph{smaller is stronger} response is observed in all cases, with graded-enhanced models predicting a stiffer response relative to the conventional plasticity case ($\ell=0$). Moreover, there is a strong sensitivity of the results to the choice of length scale parameter, with $\ell_{MIN}$ approaching the conventional plasticity solution. Second, finite element computations show that larger bending moments are attained, for the same applied curvature, with increasing $k$. And third, the $\ell_{FGM}$ case shows larger differences with conventional plasticity as $k$ decreases. The different effects observed with varying $k$ can be understood from the sensitivity of the length scale parameter (smaller $k$ entails larger values of $\ell_e$) and the stiffness (the average volume fraction of Ti increases with $k$) to the material gradient index. Size effects can be neglected in Ti-Al functionally graded foils with thicknesses on the order of 10 $\mu$m if $k$ is larger than 1.\\

\begin{figure}[H] 
    \centering
    \includegraphics[scale=1.1]{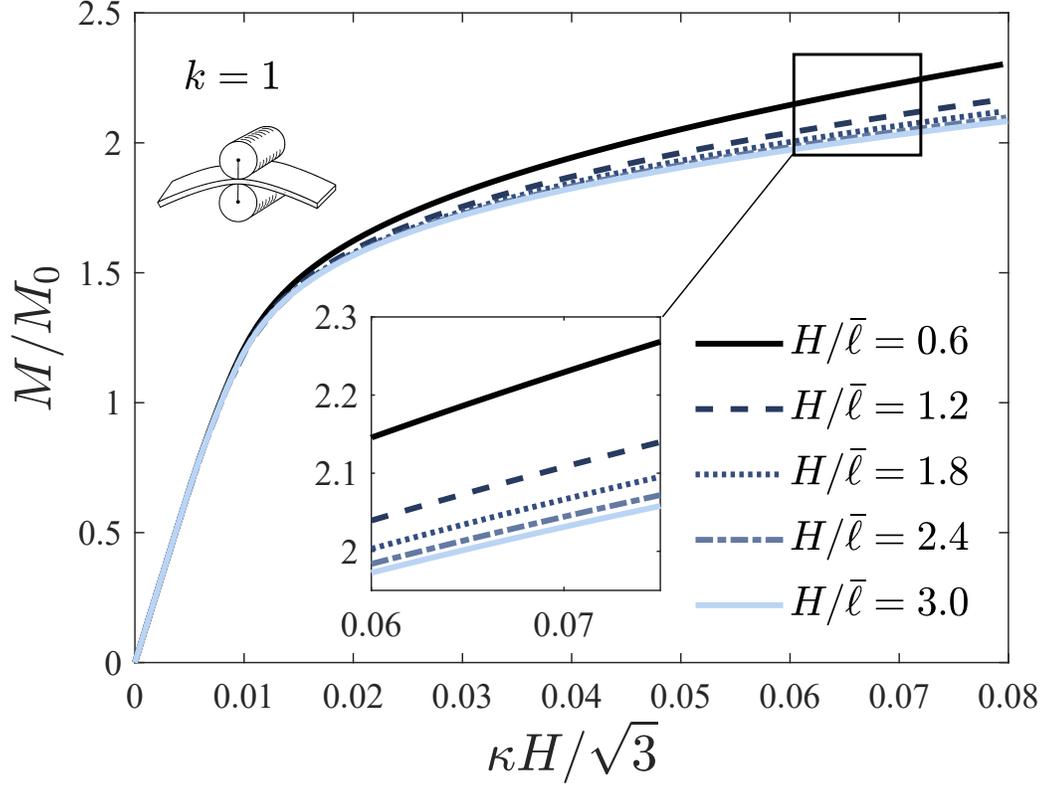}
    \caption{Normalized moment versus curvature for different foil thicknesses in a functionally graded solid.}
    \label{fig:MvskappaL}
\end{figure}

Strain gradient hardening becomes increasingly noticeable for thinner foils, in agreement with expectations. Fig. \ref{fig:MvskappaL} shows the bending response for different foil thickness, with $H$ normalized by the average length scale $\bar{\ell}=3.285$ $\mu$m. The case of a gradually varying $\ell$ is considered and the material gradient index is assumed to be $k=1$. The applied curvature is normalized in all cases by the reference thickness of the beam (i.e., $H=0.01$ mm, as considered in Fig. \ref{fig:Mvskappa}). Results reveal that thinner specimens strain harden more than thicker ones, as observed in micro-bending experiments on homogeneous materials \cite{Stolken1998}.

\begin{figure}[H] 
    \centering
    \includegraphics[scale=1.1]{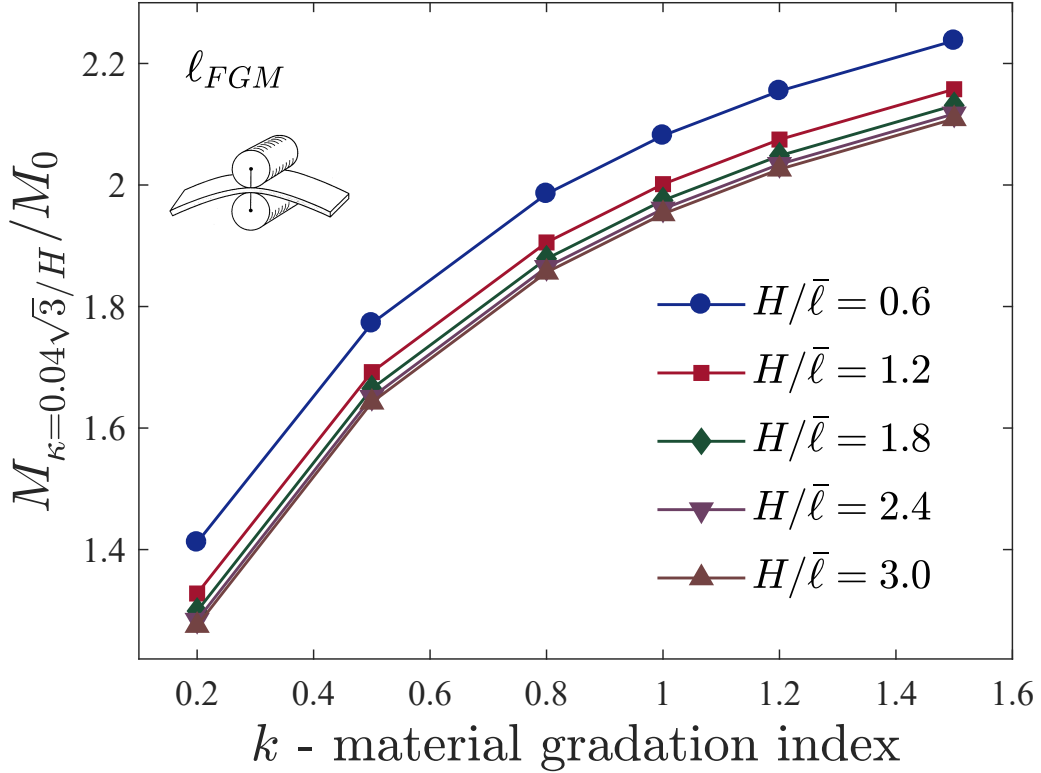}
    \caption{Normalized bending moment in a functionally graded solid at a curvature $\kappa=0.04 \sqrt{3} / H$ for different foil thicknesses and as a function of the material gradient index $k$.}
    \label{fig:Mvskappak}
\end{figure}

Next, we analyze the influence of the material gradient index in the moment versus curvature response in Fig. \ref{fig:Mvskappak}. The maximum value of the bending moment is computed for a curvature level of $\kappa H / \sqrt{3}=0.04$ for different values of $k$ and different thicknesses. As shown in the figure, the bending moment increases with decreasing $H$ and increasing $k$. This could be expected as gradient effects become relevant when the thickness approaches the length scale parameter and increasing $k$ augments the volume fraction of the stiffer constituent. However, one should note that larger values of the material gradation index make $\ell_e$ decrease, lessening strain gradient effects (i.e., for sufficiently thin films, GNDs may dominate the response and inverse the qualitative trend observed). Notwithstanding, results show that, for foil thicknesses characteristic of microelectronic applications, (i) decreasing $k$ makes the response softer, and (ii) size effects are relevant and strain gradient theories are needed to model the mechanical response.

\subsection{Indentation of FGM specimens}
\label{Sec:Indentation}

\begin{figure}[H]
\makebox[\linewidth][c]{%
        \begin{subfigure}[b]{0.6\textwidth}
                \centering
                \includegraphics[scale=0.5]{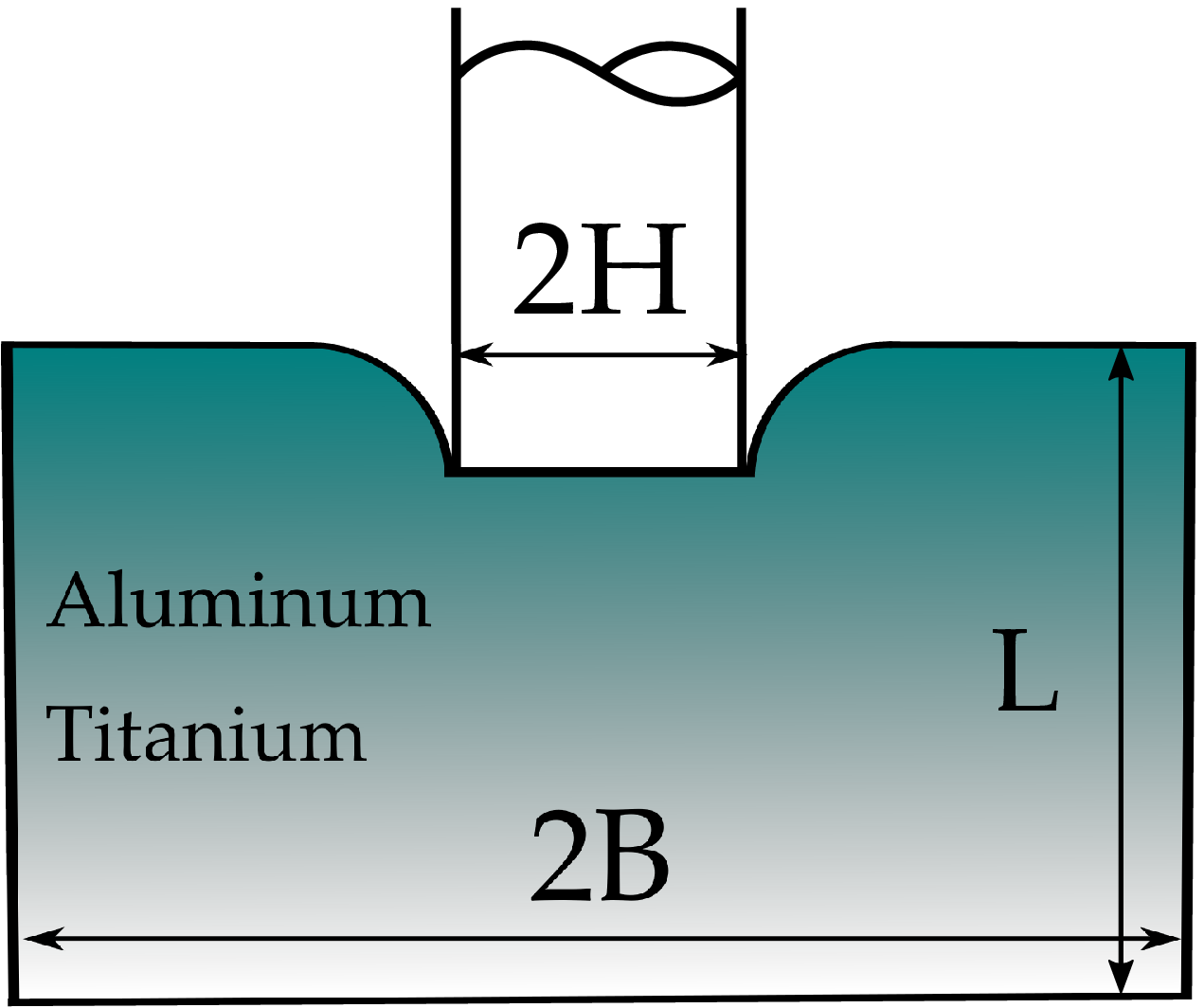}
                \caption{}
                \label{fig:IndentGeom}
        \end{subfigure}
        \begin{subfigure}[b]{0.6\textwidth}
                \centering
                \includegraphics[scale=0.5]{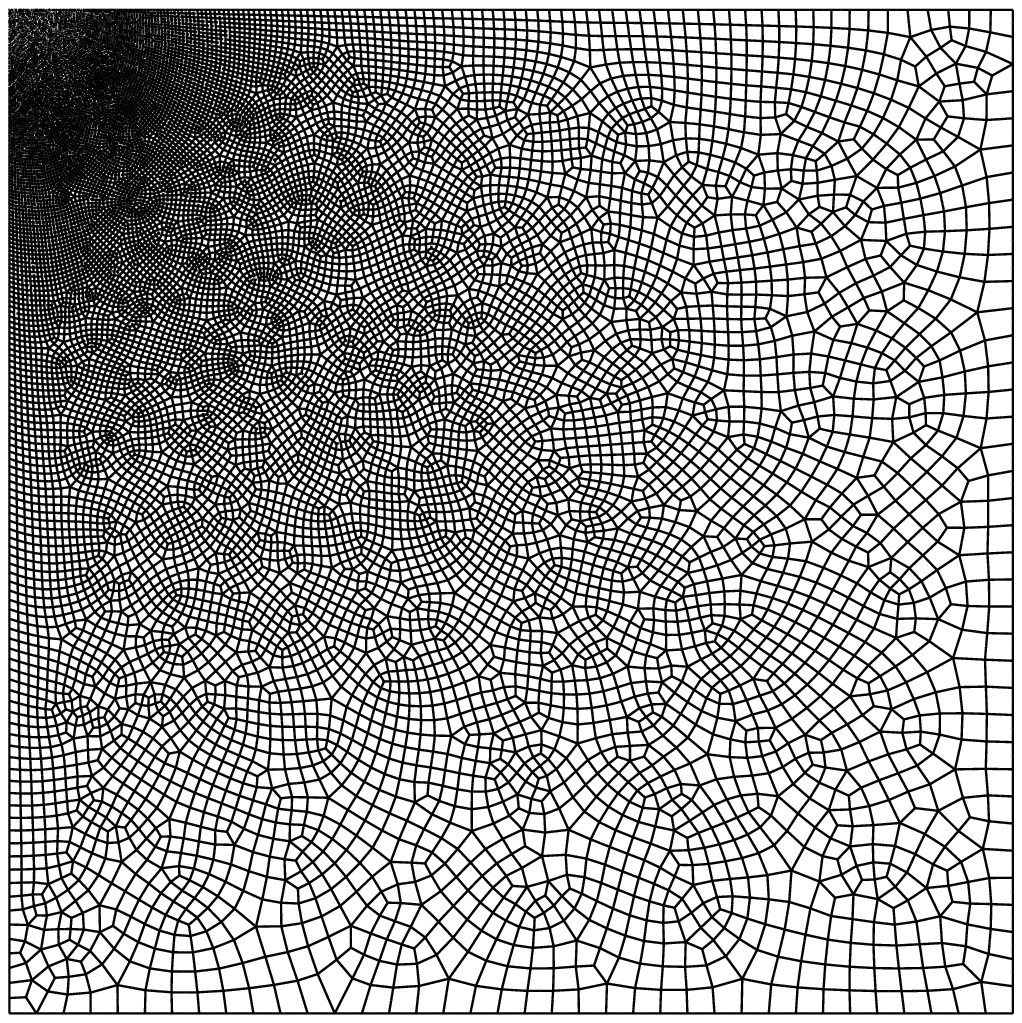}
                \caption{}
                \label{fig:IndentMesh}
        \end{subfigure}}     
        \caption{Indentation in elasto-plastic FMGs: (a) geometry, and (b) finite element mesh.}\label{fig:Indent0}
\end{figure}

Since the pioneering works by Poole et al. \cite{Poole1996} and Nix and Gao \cite{Nix1998}, micro- and nano-indentation experiments have been widely used to characterize size effects in metals, as well as to develop and benchmark gradient plasticity formulations. We examine the size-dependent indentation response of FGMs modeling a flat indenter under perfect friction conditions (see, e.g., \cite{Panteghini2016}). Loading is imposed by prescribing the displacements of the indented surface and material properties are assumed to vary gradually in the vertical direction from aluminum (top) to titanium (bottom) - see Fig. \ref{fig:IndentGeom}. Unless otherwise stated, the indenter size is chosen to be $H=0.02$ mm and the characteristic length of the indented solid is given by $B/H=25$. We take advantage of symmetry in the finite element model and choose to examine a square specimen of side length $B$ (i.e., $L=B$). As shown in Fig. \ref{fig:IndentMesh} the mesh is particularly refined in the vicinity of the indented surface, where the vertical displacement $U$ is prescribed. The horizontal displacement is set to zero to reproduce perfect friction conditions. We assume plane strain conditions and employ a total of 18565 quadratic quadrilateral elements. 

\begin{figure}[H]
\makebox[\linewidth][c]{%
        \begin{subfigure}[b]{0.6\textwidth}
                \centering
                \includegraphics[scale=0.65]{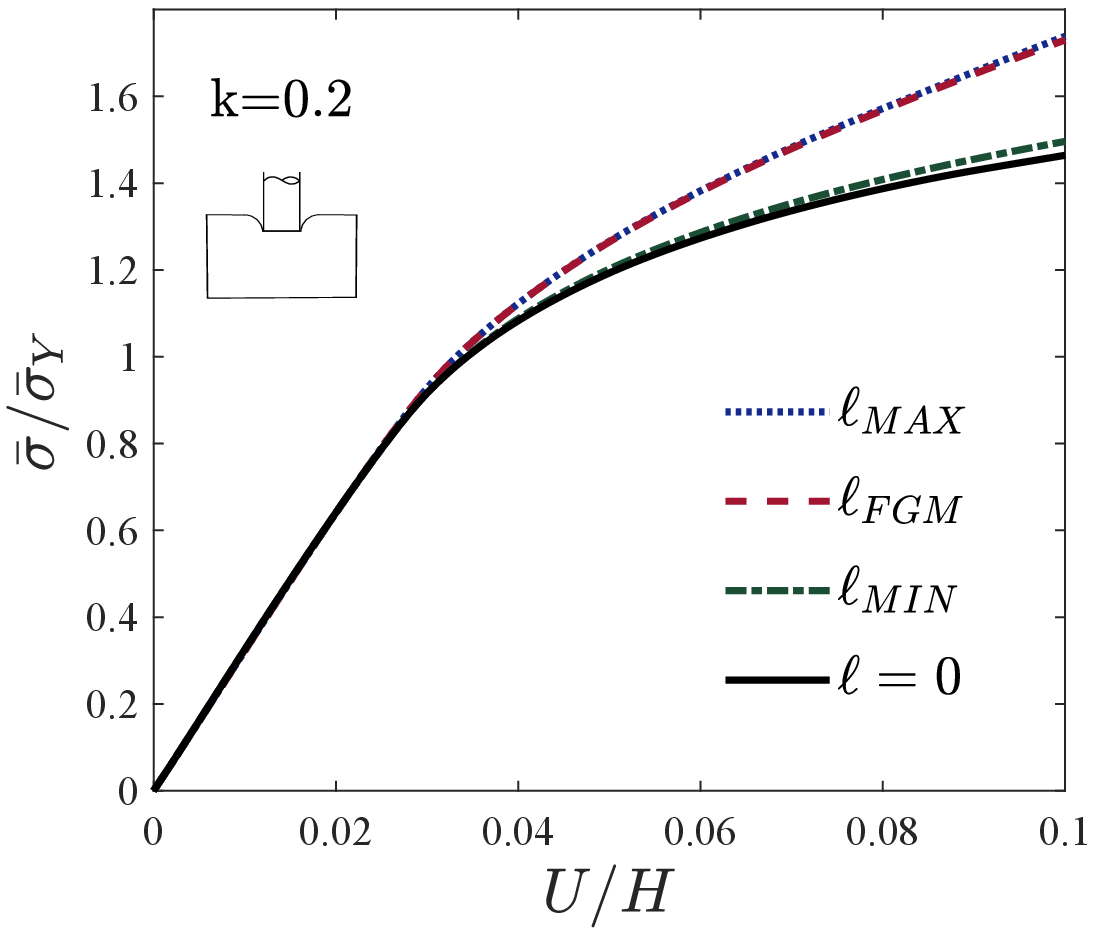}
                \caption{}
                \label{fig:Indent1k02}
        \end{subfigure}
        \begin{subfigure}[b]{0.6\textwidth}
                \raggedleft
                \includegraphics[scale=0.65]{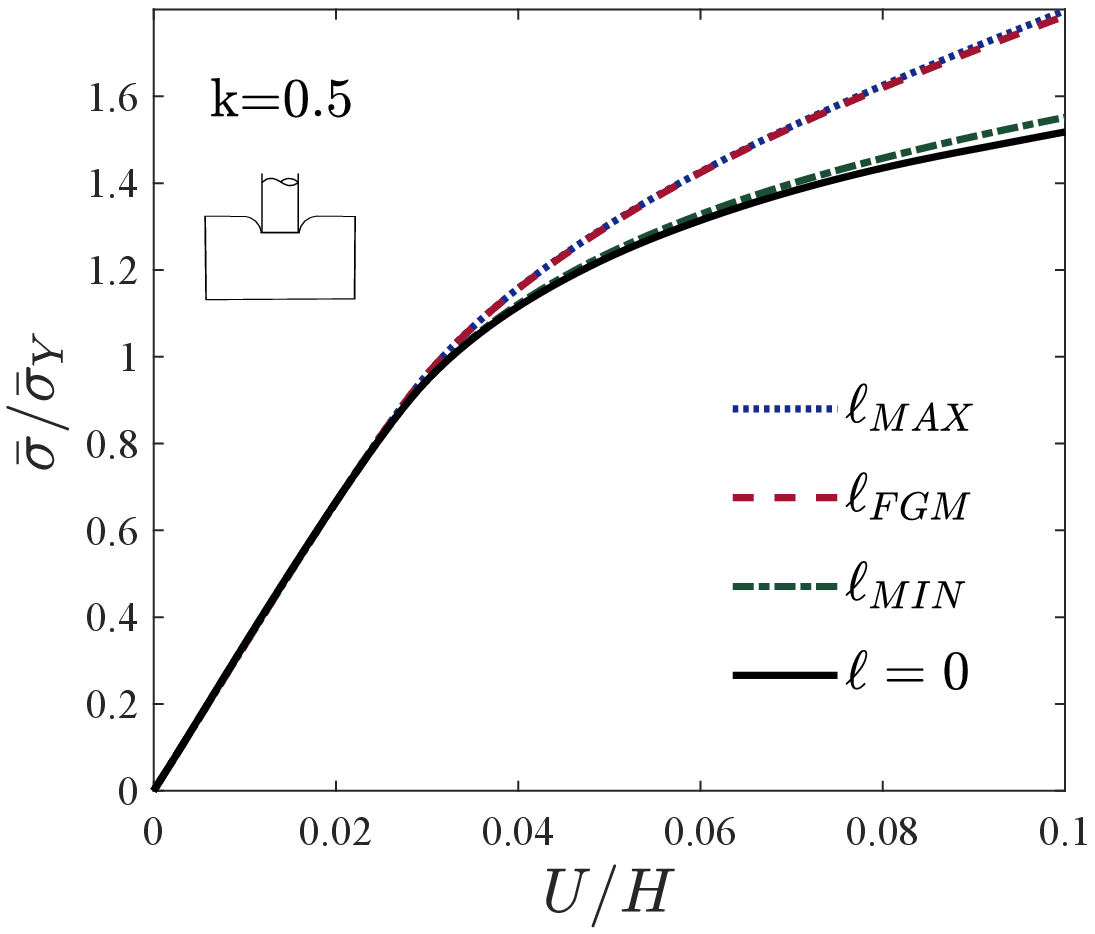}
                \caption{}
                \label{fig:Indent1k05}
        \end{subfigure}}

\makebox[\linewidth][c]{%       
        \begin{subfigure}[b]{0.6\textwidth}
                \centering
                \includegraphics[scale=0.65]{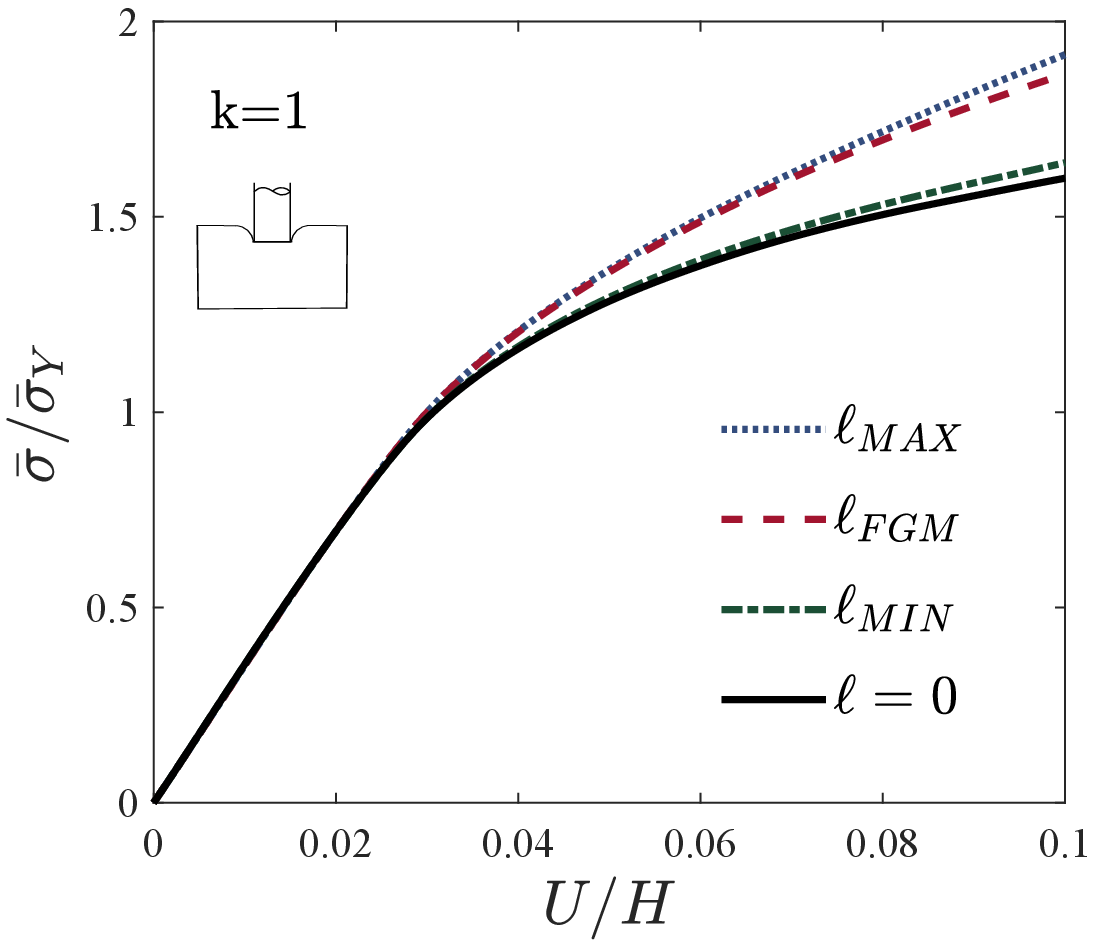}
                \caption{}
                \label{fig:Indent1k1}
        \end{subfigure}
        \begin{subfigure}[b]{0.6\textwidth}
                \raggedleft
                \includegraphics[scale=0.65]{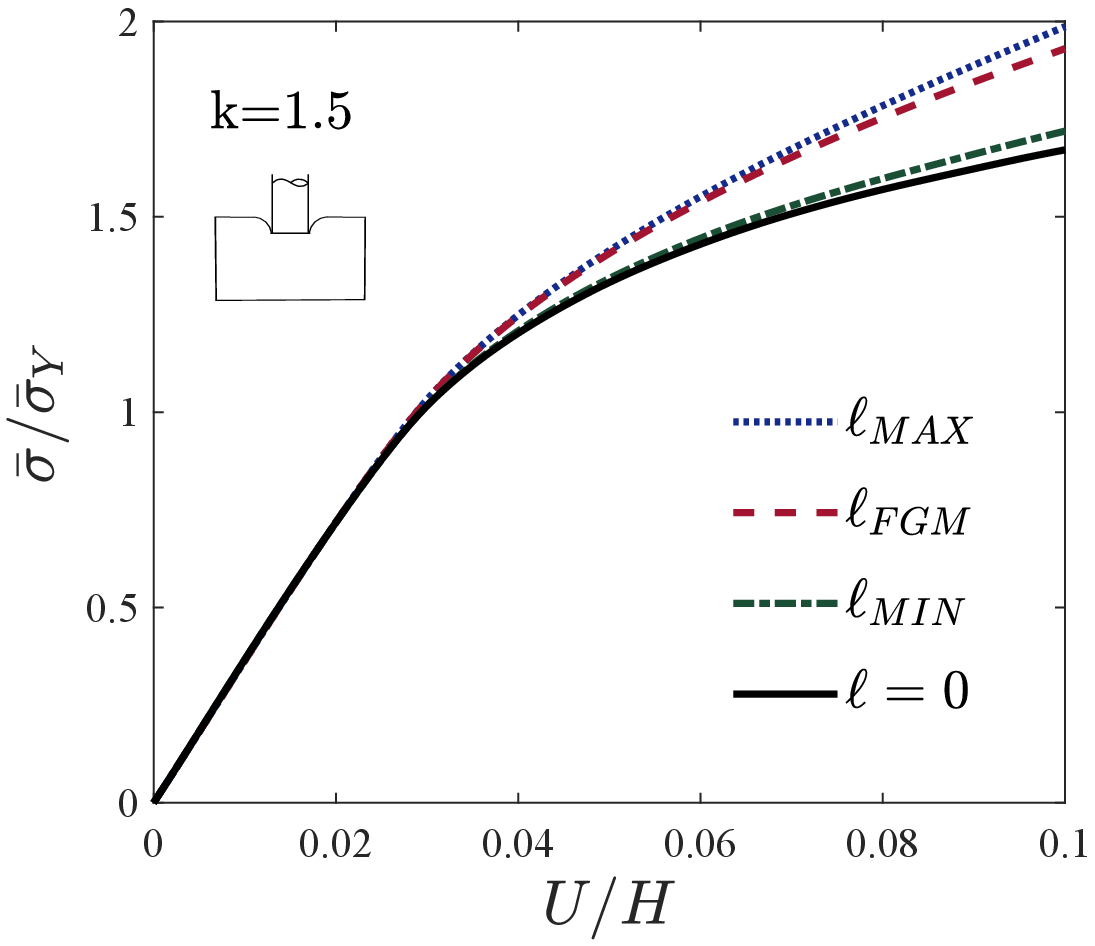}
                \caption{}
                \label{fig:Indent1k15}
        \end{subfigure}
        }       
        \caption{Average non-dimensional indentation pressure as a function of non-dimensional indentation depth for conventional ($\ell=0$) and strain gradient plasticity theories. A functionally graded length parameter ($\ell_{FGM}$) has been considered, along with the limit cases of the stiffer ($\ell_{MAX}$) and the softer ($\ell_{MIN}$) compounds. Material gradient indexes: (a) $k=0.2$, (b) $k=0.5$, (c) $k=1$, and (d) $k=1.5$.}\label{fig:Indent1}
\end{figure}

\begin{figure}[H] 
    \centering
    \includegraphics[scale=1.2]{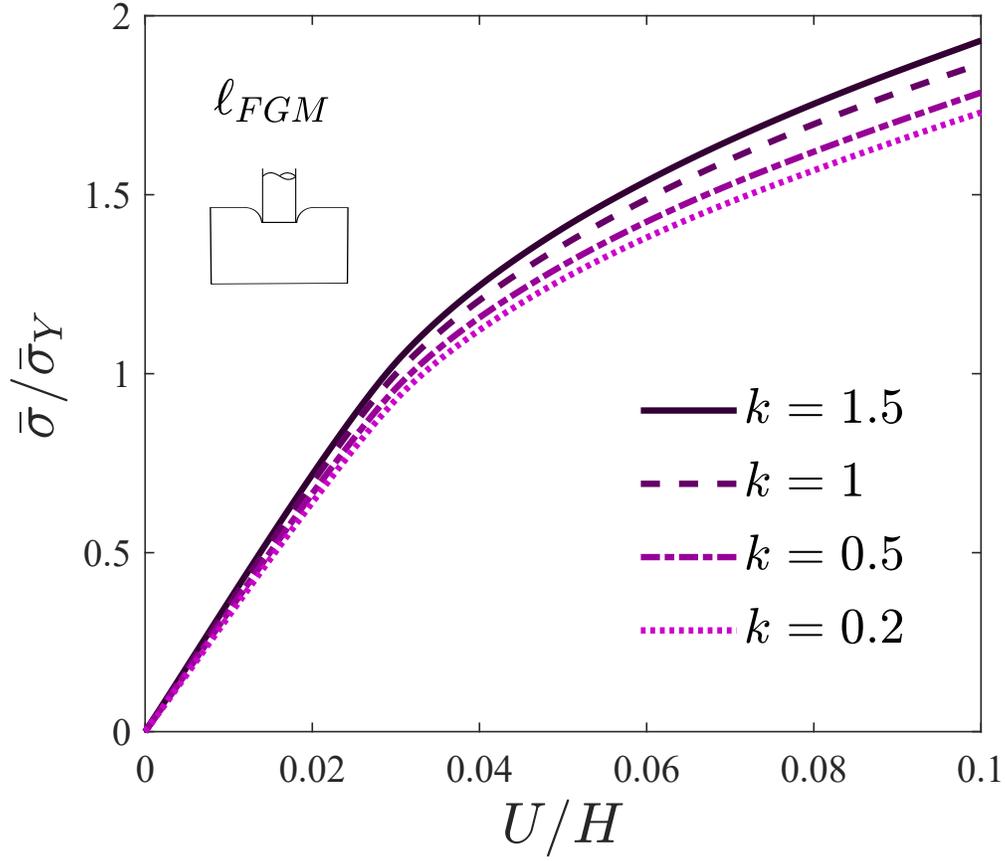}
    \caption{Average non-dimensional indentation pressure as a function of non-dimensional indentation depth for different material gradient indexes in a functionally graded strain gradient solid.}
    \label{fig:Indent2}
\end{figure}

The average normalized indentation pressure $\bar{\sigma}/\bar{\sigma}_Y$ is given as a function of the normalized indentation depth $U/H$ in Fig. \ref{fig:Indent1} for various choices of the material gradient index $k$. Here, $\bar{\sigma}$ is the average normal stress component on the indented surface while $\bar{\sigma}_Y$ denotes the average yield stress of the two constituents (578 MPa). Results show that the response for a functionally graded length parameter follows closely that of $\ell_{MAX}$. This is traced to the fact that we are indenting the aluminum region and consequently $\ell_{FGM}$ adopts values comparable to its maximum within the plastic zone. On the other hand, $\ell_{MIN}$ predictions only slightly elevate the indentation pressure relative to the conventional plasticity case for an indenter size of 10 $\mu$m. In addition, larger values of the material gradient index entail larger differences between $\ell_{FGM}$ and $\ell_{MAX}$ predictions; higher values of $k$ result in a stronger reduction of the length parameter with distance to the indented surface. It is also observed that larger values of the gradient index lead to stiffer responses as the average volume fraction of Ti increases; this is further examined in Fig. \ref{fig:Indent2}. 

Fig. \ref{fig:Indent2} shows the dependence of the indentation response on the material gradient index $k$ for the $\ell_{FGM}$ case. Higher values of the volume fraction exponent lead to a stiffer response; increasing $k$ brings higher values of $\ell_e$ in the plastic zone but also decreases the average volume fraction of the softer material. The indentation response is less sensitive to changes in the material gradation profile when the effect of dislocation hardening is accounted for.

\begin{figure}[H] 
    \centering
    \includegraphics[scale=1.2]{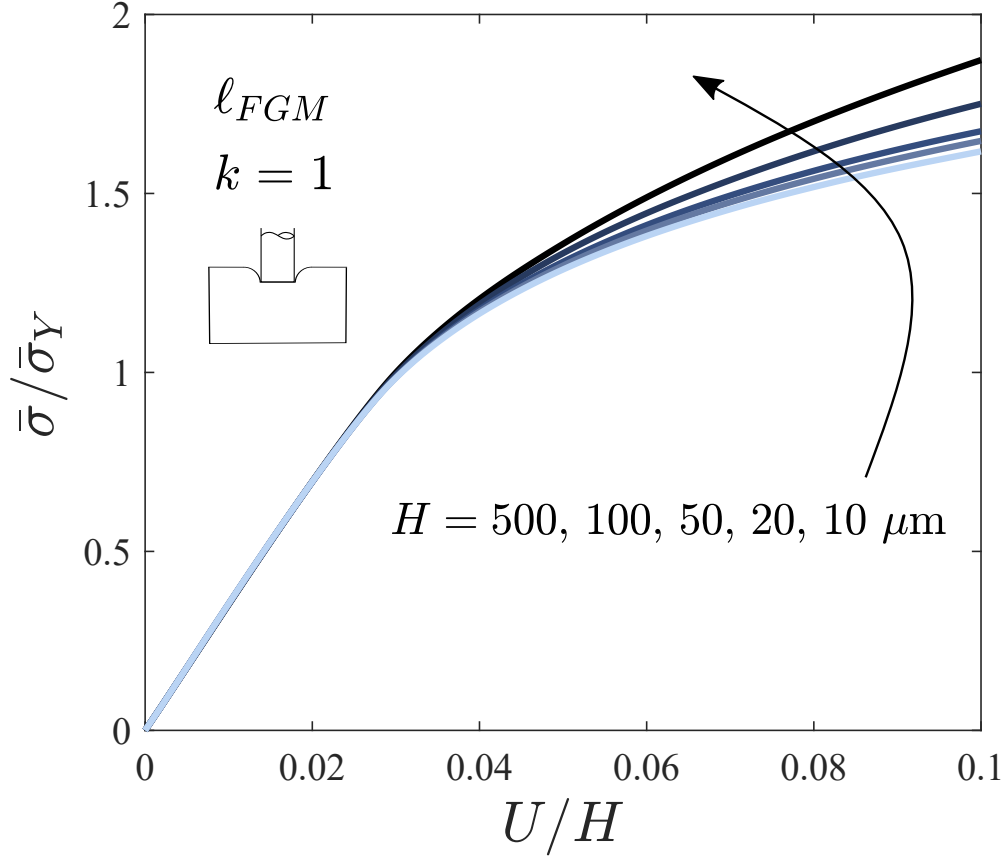}
    \caption{Average non-dimensional indentation pressure as a function of non-dimensional indentation depth for different indenter sizes in a functionally graded strain gradient solid. The indentation depth is normalized by the reference indenter size: $H=20$ $\mu$m.}
    \label{fig:Indent3}
\end{figure}

Finally, we investigate the sensitivity of the indentation pressure to the indenter size. As shown in Fig. \ref{fig:Indent3}, gradient-enriched results show an increase in strain hardening with diminishing size. Strong size effects are observed in small scale indentation of functionally graded solids when indenting on the softer edge. 

\subsection{Fracture mechanics in elasto-plastic FGM specimens}
\label{Sec:Fracture}

Geometrically necessary dislocations play a fundamental role in fracture problems as, independently of the size of the specimen, the plastic zone adjacent to the crack tip is physically small and contains strong spatial gradients of deformation. These large dislocation densities promote strain hardening and lead to high crack tip stresses that conventional plasticity is unable to capture \cite{IJP2016,CM2017}. Enriching continuum theories to accurately compute stress distributions has proven to be pivotal in hydrogen embrittlement,  where crack tip stresses govern hydrogen diffusion \cite{IJHE2016} and decohesion \cite{AM2016}. Fracture in FGMs has been investigated extensively in the context of conventional continuum theories of elasticity and plasticity (see, e.g. \cite{Tvergaard2002,Batra2005,IJMMD2015,Ooi2015} and references therein). We use the present mechanism-based FGM formulation to investigate the influence of plastic strain gradients ahead of cracks in functionally graded solids.\\

\begin{figure}[H]
\centering
\begin{subfigure}[h]{0.99\textwidth}
\centering
\begin{tikzpicture}[thick]
\shade[top color=teal,bottom color=white, draw=black] (0,0) rectangle (10,2.5);
\draw[thick,->] (5,1.3) -- (5.7,1.3) node[anchor=north west] {x axis};
\draw[thick,->] (5,1.3) -- (5,2) node[anchor=south east] {y axis};
\foreach \y in {0,0.5,1,1.5,2,2.5}
    \draw[black,thick,->] (0,\y) -- (-0.5,\y);
%\draw (0,0) -- (-0.4,0.4) --(-0.4,-0.4) -- cycle;
\foreach \y in {0,0.5,1,1.5,2,2.5}
	\draw[black,thick,->] (10,\y) -- (10.5,\y);
\draw[black,thick] (5,0) -- (5,0.833);
\node at (2,2) {Aluminum};
\node at (2,0.5) {Titanium};
\draw[black,very thick,<->] (0,-0.5) -- (10,-0.5); 
\draw[black,very thick,-] (0,-0.75) -- (0,-0.25);
\draw[black,very thick,-] (10,-0.75) -- (10,-0.25);
\node at (5,-0.75) {W};
\draw[black,very thick,<->] (-1,0) -- (-1,2.5); 
\draw[black,very thick,-] (-1.25,0) -- (-0.75,0);
\draw[black,very thick,-] (-1.25,2.5) -- (-0.75,2.5);
\node at (-1.3,1.25) {H};
\draw[black,very thick,<->] (5.25,0) -- (5.25,0.833); 
\node at (5.5,0.5) {a};
\end{tikzpicture}
\caption{}
\label{fig:FractureGeometry}
\end{subfigure}
\vspace{20pt}
\begin{subfigure}[h]{0.99\textwidth}
\centering
\includegraphics[scale=0.65]{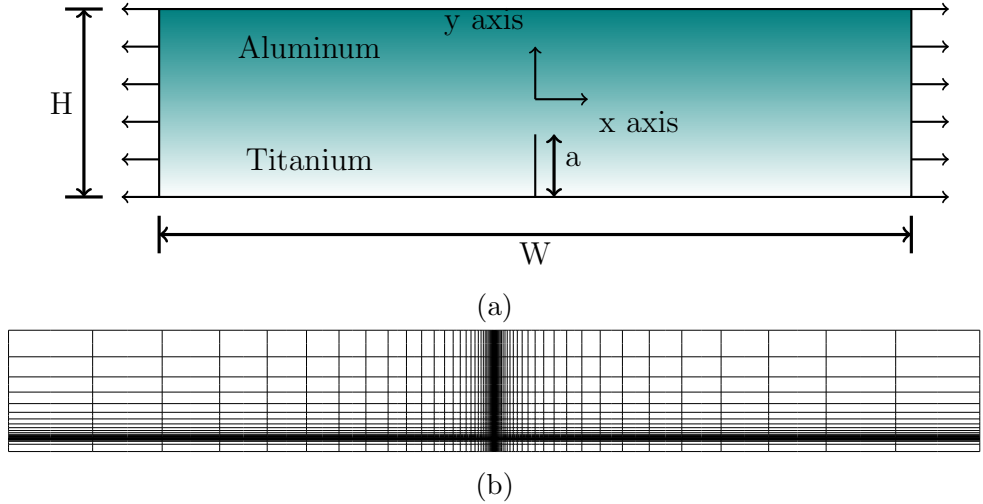}
\caption{}
\label{fig:FractureMesh}
\end{subfigure}
       
        \caption{Fracture analysis of a functionally graded plate: (a) Geometry and loading configuration and (b) finite element mesh.}\label{fig:FractureGeoMesh}
\end{figure}

Crack tip fields are computed in an edge-cracked plate of height-to-width ratio $W/H=8$ and crack size $a/H=0.1$. As shown in Fig. \ref{fig:FractureGeometry}, the crack starts in the titanium edge and the plate is loaded in mode I by prescribing the displacements at horizontal sides. The stress intensity factor reads $K_I= Y \sigma_R \sqrt{\pi a}$, where $Y$ is the geometry factor and the remote stress is given by $\sigma_R=\sigma_{xx} \left( x=W/2 \right)$. A reference length of the plastic zone size is defined based on Irwin,
\begin{equation}
R_p = \frac{1}{3 \pi} \left( \frac{K_0}{\bar{\sigma}_Y} \right)^2
\end{equation}

\noindent with $\bar{\sigma}_Y$ being the average yield stress. After a sensitivity study, the specimen is discretized by means of 7560 finite elements. We use the same type of element as for the indentation and bending case studies, 8-node quadrilateral plane strain with reduced integration. As shown in Fig. \ref{fig:FractureMesh}, the mesh is gradually refined in the vicinity of the crack; the use of a very fine mesh close to the crack tip is essential to capture the effect of steep plastic strain gradients and ensure convergence of the solution.\\

\begin{figure}[H] 
    \centering
    \includegraphics[scale=1.2]{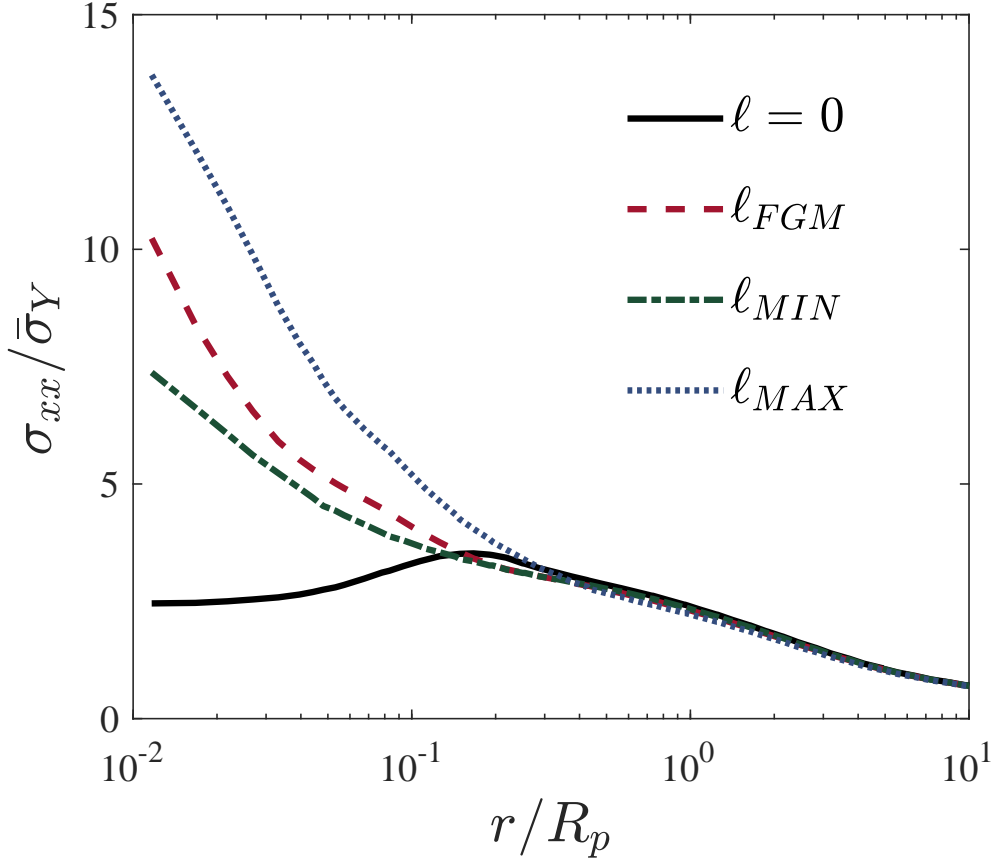}
    \caption{Crack opening stress distribution in an FGM specimen for conventional ($\ell=0$) and strain gradient plasticity theories. A functionally graded length parameter ($\ell_{FGM}$) has been considered, along with the limit cases of the stiffer ($\ell_{MAX}$) and the softer ($\ell_{MIN}$) compounds. Material gradient index: $k=1$.}
    \label{fig:CrackTipS11}
\end{figure}

Crack opening stresses are shown in Fig. \ref{fig:CrackTipS11} for a given material gradient index ($k=1$) as a function of the distance to the crack tip $r$. The stress values are normalized by the average yield stress $\bar{\sigma}_Y$ while the horizontal axis shows the ratio $r/R_p$ in logarithmic scale. As in the bending and indentation cases, results are computed for four values of the length scale parameter: (i) $\ell=0$ (conventional plasticity), (ii) $\ell_{MAX}$ (Ti), (iii) $\ell_{MIN}$ (Al), and (iv) variable $\ell_{FGM}$ (\ref{Eq:length}). Fig. \ref{fig:CrackTipS11} shows that all cases agree far away from the crack tip but differences become significant within a fraction of the plastic zone size. Large gradients of plastic strain in the vicinity of the crack promote local hardening and elevate opening stresses. The conventional plasticity solution reaches a peak at a certain distance from the crack tip and then decreases as stresses become influenced by large strains and crack blunting. This stress triaxiality reduction is not observed in any of the $\ell \neq 0$ scenarios, which translates into significant differences in the magnitude of the opening stress predicted at the crack tip. Specifically, $\sigma_{xx}$ is roughly 3, 4 and 5.5 times the conventional plasticity result in the vicinity of the crack for $\ell_{MIN}$, $\ell_{FGM}$ and $\ell_{MAX}$, respectively. As it could be expected, by further reducing $\ell$ beyond $\ell_{MIN}$ the solution transitions towards the conventional plasticity prediction ($\ell=0$).

\begin{figure}[H] 
    \centering
    \includegraphics[scale=0.8]{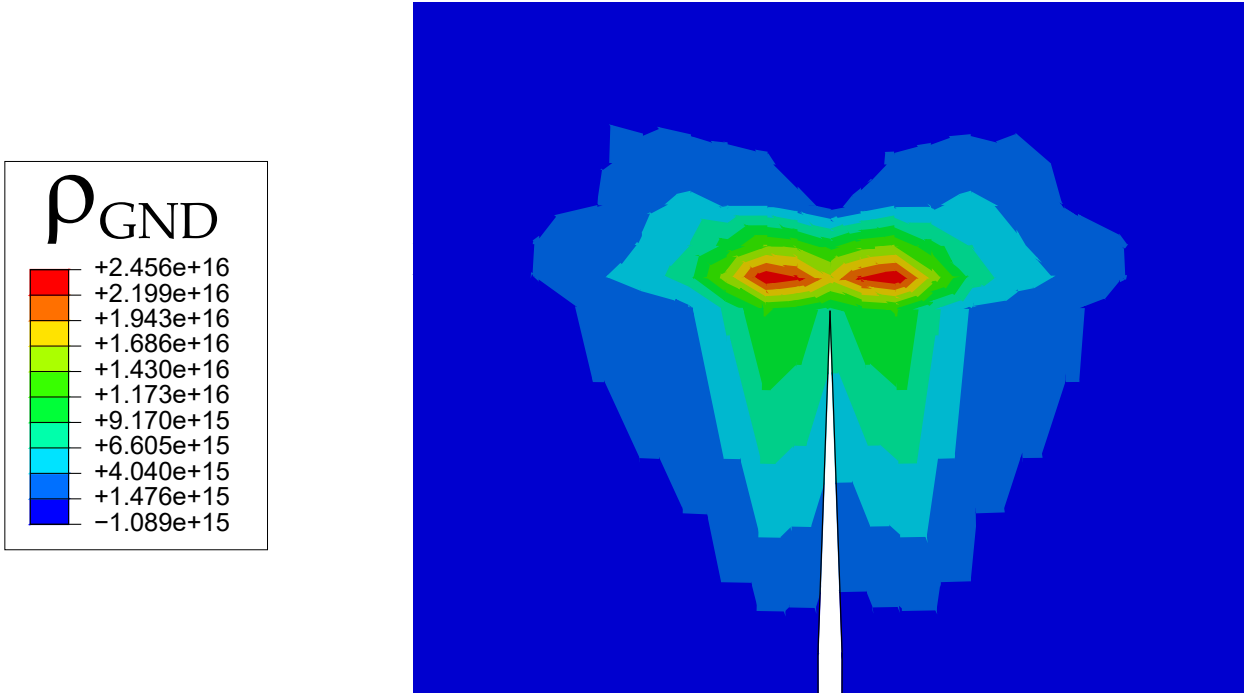}
    \caption{Contour plot of the density of geometrically necessary dislocations in m$^{-2}$. Results are shown for the case of $\ell_{MIN}$ and $R_p=0.00027$ mm.}
    \label{fig:GNDs}
\end{figure}

The stress attenuation intrinsic to conventional plasticity is not observed in the graded-enhanced results. This is due to the contribution of the strain gradients to material work hardening, which significantly lowers crack tip blunting. Hardening increases in the vicinity of the crack due to the large GND density; geometrically necessary dislocations act as obstacles to the motion of statistically stored dislocations, reducing plastic deformation and elevating the stresses. Fig. \ref{fig:GNDs} shows the GND density contours predicted by the present formulation - see Eq. (\ref{Eq:rhoG}). Results are shown for the length parameter of Al, taking the Burgers vector to be that of fcc materials. Large GND densities are obtained ahead of the crack, dominating the contribution to the flow stress.

\begin{figure}[H] 
    \centering
    \includegraphics[scale=1.1]{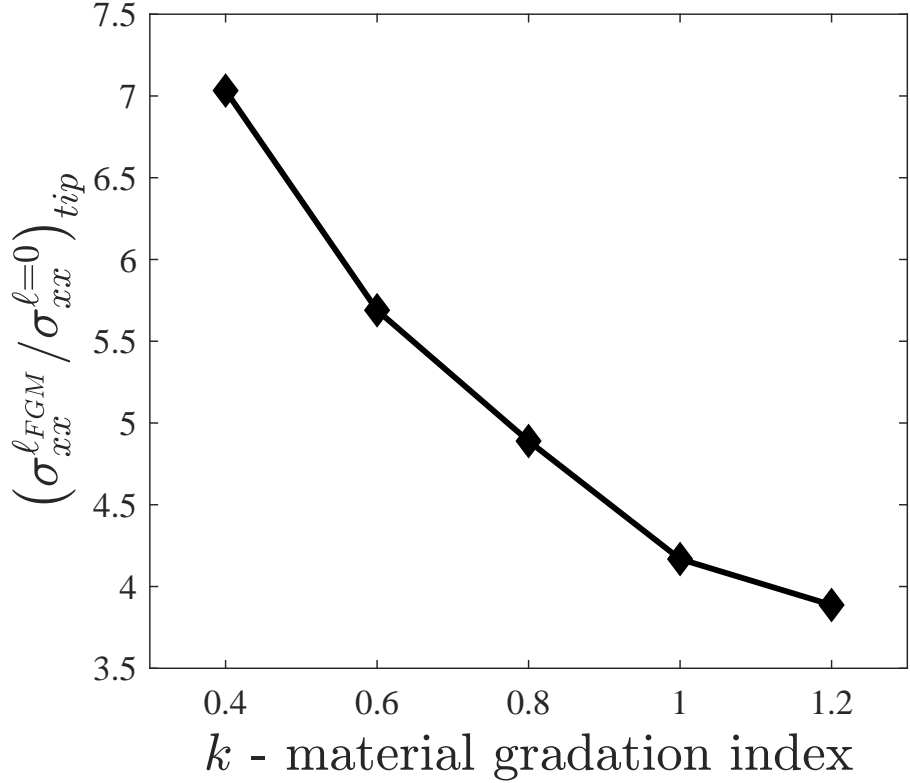}
    \caption{Crack tip stress ratio as a function of the material gradient index $k$. Results are shown for the case of $\ell_{FGM}$ and $R_p=0.0027$ mm.}
    \label{fig:kCrack}
\end{figure}

We investigate the influence of the material gradation profile on the stress elevation intrinsic to gradient dislocation hardening. Fig. \ref{fig:kCrack} shows the opening stress ratio between conventional and strain gradient plasticity predictions as a function of the material gradient index $k$. The $\ell \neq 0$ solution corresponds to that obtained considering a functionally graded profile of the length scale parameter.

Results reveal a significant increase in the gradient-enhanced stress elevation with decreasing $k$. As shown in Fig. \ref{fig:LengthScaleVar}, high material gradation indexes translate into convex aluminum volume fraction profiles along the relevant dimension of the specimen, with the effective length parameter adopting lower values over the characteristic length of the specimen. Thus, $\ell_e$ in the vicinity of the crack will increase with decreasing $k$ as the softer compound gains weight. The stress elevation can therefore be substantially attenuated if the gradation profile of Ti is concave downward.

\section{Conclusions}
\label{Sec:Concluding remarks}

Size effects in elastic-plastic functionally graded materials (FGMs) have been extensively investigated by means of a strain gradient plasticity formulation for graded materials. A suitable homogenization scheme is employed to accurately sample material properties as a function of the volume fraction of material. The analysis of (i) bending of thin functionally graded foils, (ii) small scale indentation in the gradient direction, and (iii) stress fields in cracked FGM specimens reveals that,

\begin{itemize}
  \item Gradients of plastic strain have a profound effect on the mechanical response of metal-based FGMs. Large geometrically necessary dislocation densities significantly elevate the flow strength when non-homogeneous plastic deformation is confined to small volumes.
  \item The \emph{smaller is stronger} behavior is very sensitive to the material gradation, with aluminum convex volume fraction profiles exhibiting considerably smaller size effects. The design strategy of increasing the volume fraction of the harder metal to maximize stiffness will be compromised when the specimen size is on the order of a micron (or less).
  \item Neglecting the length scale variation across the specimen brings important differences in the macroscopic response. The intrinsic gradation of the length scale parameter must be accounted for to accurately characterize gradient effects in FGMs. 
\end{itemize}

Results highlight the need to conduct critical experiments to accurately characterize the behaviour of elastic-plastic FGMs at the micro scale.
 
\section{Acknowledgements}
\label{Acknowledge of funding}

E. Mart\'{\i}nez-Pa\~neda acknowledges financial support from the Ministry of Economy and Competitiveness of Spain through grant MAT2014-58738-C3 and the People Programme (Marie Curie Actions) of the European Union's Seventh Framework Programme (FP7/2007-2013) under REA grant agreement n$^{\circ}$ 609405 (COFUNDPostdocDTU).

%% The Appendices part is started with the command \appendix;
%% appendix sections are then done as normal sections

%% If you have bibdatabase file and want bibtex to generate the
%% bibitems, please use
%%
%%  \bibliographystyle{elsarticle-harv} 
%%  \bibliography{<your bibdatabase>}

%% else use the following coding to input the bibitems directly in the
%% TeX file.

\bibliographystyle{elsarticle-num}
\bibliography{library}

\begin{thebibliography}{10}
\expandafter\ifx\csname url\endcsname\relax
  \def\url#1{\texttt{#1}}\fi
\expandafter\ifx\csname urlprefix\endcsname\relax\def\urlprefix{URL }\fi
\expandafter\ifx\csname href\endcsname\relax
  \def\href#1#2{#2} \def\path#1{#1}\fi

\bibitem{Fu2004}
Y.~Fu, H.~Du, W.~Huang, S.~Zhang, M.~Hu, {TiNi-based thin films in MEMS
  applications: A review}, Sensors and Actuators, A: Physical 112~(2-3) (2004)
  395--408.

\bibitem{Natarajan2012}
S.~Natarajan, S.~Chakraborty, M.~Thangavel, S.~Bordas, T.~Rabczuk,
  {Size-dependent free flexural vibration behavior of functionally graded
  nanoplates}, Computational Materials Science 65 (2012) 74--80.

\bibitem{Natarajan2014}
S.~Natarajan, {On the application of the partition of unity method for nonlocal
  response of low-dimensional structures}, Journal of the Mechanical Behavior
  of Materials 23~(5-6) (2014) 153--168.

\bibitem{Lou2016}
J.~Lou, L.~He, J.~Du, H.~Wu, {Nonlinear analyses of functionally graded
  microplates based on a general four-variable refined plate model and the
  modified couple stress theory}, Composite Structures 152 (2016) 516--527.

\bibitem{Simsek2017}
M.~Simsek, M.~Aydın, {Size-dependent forced vibration of an imperfect
  functionally graded (FG) microplate with porosities subjected to a moving
  load using the modified couple stress theory}, Composite Structures 160
  (2017) 408--421.

\bibitem{Sahmani2018}
S.~Sahmani, M.~Aghdam, T.~Rabczuk, {Nonlinear bending of functionally graded
  porous micro/nano-beams reinforced with graphene platelets based upon
  nonlocal strain gradient theory}, Composite Structures 186 (2018) 68--78.

\bibitem{Momeni2018}
S.~Momeni, M.~Asghari, {The second strain gradient functionally graded beam
  formulation}, Composite Structures 188 (2018) 15--24.

\bibitem{Tsiatas2017}
G.~C. Tsiatas, N.~G. Babouskos, {Elastic-plastic analysis of functionally
  graded bars under torsional loading}, Composite Structures 176 (2017)
  254--267.

\bibitem{Amirpour2017}
M.~Amirpour, R.~Das, S.~Bickerton, {An elasto-plastic damage model for
  functionally graded plates with in-plane material properties variation:
  Material model and numerical implementation}, Composite Structures 163 (2017)
  331--341.

\bibitem{Nix1998}
W.~D. Nix, H.~Gao, {Indentation size effects in crystalline materials: A law
  for strain gradient plasticity}, Journal of the Mechanics and Physics of
  Solids 46~(3) (1998) 411--425.

\bibitem{Fleck1994}
N.~A. Fleck, G.~M. Muller, M.~F. Ashby, J.~W. Hutchinson, {Strain gradient
  plasticity: Theory and experiment}, Acta Metallurgica et Materialia 42~(2)
  (1994) 475--487.

\bibitem{Stolken1998}
J.~St{\"{o}}lken, A.~Evans, {A microbend test method for measuring the
  plasticity length scale}, Acta Materialia 46~(14) (1998) 5109--5115.

\bibitem{Gao1999}
H.~Gao, Y.~Hang, W.~D. Nix, J.~W. Hutchinson, {Mechanism-based strain gradient
  plasticity - I. Theory}, Journal of the Mechanics and Physics of Solids
  47~(6) (1999) 1239--1263.

\bibitem{Fleck2001}
N.~A. Fleck, J.~W. Hutchinson, {A reformulation of strain gradient plasticity},
  Journal of the Mechanics and Physics of Solids 49~(10) (2001) 2245--2271.

\bibitem{Huang2004}
Y.~Huang, S.~Qu, K.~C. Hwang, M.~Li, H.~Gao, {A conventional theory of
  mechanism-based strain gradient plasticity}, International Journal of
  Plasticity 20~(4-5) (2004) 753--782.

\bibitem{Gurtin2005}
M.~E. Gurtin, L.~Anand, {A theory of strain-gradient plasticity for isotropic,
  plastically irrotational materials. Part I: Small deformations},
  International Journal of the Mechanics and Physics of Solids 53 (2005)
  1624--1649.

\bibitem{Mao2013}
Y.~Q. Mao, S.~G. Ai, D.~N. Fang, Y.~M. Fu, C.~P. Chen, {Elasto-plastic analysis
  of micro FGM beam basing on mechanism-based strain gradient plasticity
  theory}, Composite Structures 101 (2013) 168--179.

\bibitem{Hwang2003}
K.~C. Hwang, H.~Jiang, Y.~Huang, H.~Gao, {Finite deformation analysis of
  mechanism-based strain gradient plasticity: Torsion and crack tip field},
  International Journal of Plasticity 19~(2) (2003) 235--251.

\bibitem{IJSS2015}
E.~Mart{\'{i}}nez-Pa{\~{n}}eda, C.~Beteg{\'{o}}n, {Modeling damage and fracture
  within strain-gradient plasticity}, International Journal of Solids and
  Structures 59 (2015) 208--215.

\bibitem{Qu2004}
S.~Qu, Y.~Huang, H.~Jiang, C.~Liu, P.~D. Wu, K.~C. Hwang, {Fracture analysis in
  the conventional theory of mechanism-based strain gradient (CMSG)
  plasticity}, International Journal of Fracture 129~(3) (2004) 199--220.

\bibitem{TAFM2017}
E.~Mart{\'{i}}nez-Pa{\~{n}}eda, S.~del Busto, C.~Beteg{\'{o}}n, {Non-local
  plasticity effects on notch fracture mechanics}, Theoretical and Applied
  Fracture Mechanics 92 (2017) 276--287.

\bibitem{Hughes1980a}
T.~J.~R. Hughes, J.~Winget, {Finite rotation effects in numerical integration
  of rate constitutive equations arising in large-deformation analysis},
  International Journal for Numerical Methods in Engineering 15~(12) (1980)
  1862--1867.

\bibitem{AES2017}
G.~Papazafeiropoulos, M.~Mu{\~{n}}iz-Calvente, E.~Mart{\'{i}}nez-Pa{\~{n}}eda,
  {Abaqus2Matlab: A suitable tool for finite element post-processing}, Advances
  in Engineering Software 105 (2017) 9--16.

\bibitem{Mortensen1995}
A.~Mortensen, S.~Suresh, {Functionally graded metals and metal-ceramic
  composites: Part 1 Processing}, International Materials Reviews 40~(6) (1995)
  239--265.

\bibitem{Xiong2000}
H.~P. Xiong, Q.~Shen, J.~G. Li, L.~M. Zhang, R.~Z. Yuan, {Design and
  microstructures of Ti/TiAl/Al system functionally graded material}, Journal
  of Materials Science Letters 19 (2000) 989--993.

\bibitem{Idiart2009}
M.~I. Idiart, V.~S. Deshpande, N.~A. Fleck, J.~R. Willis, {Size effects in the
  bending of thin foils}, International Journal of Engineering Science
  47~(11-12) (2009) 1251--1264.

\bibitem{IJSS2016}
E.~Mart{\'{i}}nez-Pa{\~{n}}eda, C.~F. Niordson, L.~Bardella, {A finite element
  framework for distortion gradient plasticity with applications to bending of
  thin foils}, International Journal of Solids and Structures 96 (2016)
  288--299.

\bibitem{Poole1996}
W.~J. Poole, M.~F. Ashby, N.~A. Fleck, {Micro-hardness of annealed and
  work-hardened copper polycrystals}, Scripta Materialia 34~(4) (1996)
  559--564.

\bibitem{Panteghini2016}
A.~Panteghini, L.~Bardella, {On the Finite Element implementation of
  higher-order gradient plasticity, with focus on theories based on plastic
  distortion incompatibility}, Computer Methods in Applied Mechanics and
  Engineering 310 (2016) 840--865.

\bibitem{IJP2016}
E.~Mart{\'{i}}nez-Pa{\~{n}}eda, C.~F. Niordson, {On fracture in finite strain
  gradient plasticity}, International Journal of Plasticity 80 (2016) 154--167.

\bibitem{CM2017}
E.~Mart{\'{i}}nez-Pa{\~{n}}eda, S.~Natarajan, S.~Bordas, {Gradient plasticity
  crack tip characterization by means of the extended finite element method},
  Computational Mechanics 59~(5) (2017) 831--842.

\bibitem{IJHE2016}
E.~Mart{\'{i}}nez-Pa{\~{n}}eda, S.~del Busto, C.~F. Niordson, C.~Beteg{\'{o}}n,
  {Strain gradient plasticity modeling of hydrogen diffusion to the crack tip},
  International Journal of Hydrogen Energy 41~(24) (2016) 10265--10274.

\bibitem{AM2016}
E.~Mart{\'{i}}nez-Pa{\~{n}}eda, C.~F. Niordson, R.~P. Gangloff, {Strain
  gradient plasticity-based modeling of hydrogen environment assisted
  cracking}, Acta Materialia 117 (2016) 321--332.

\bibitem{Tvergaard2002}
V.~Tvergaard, {Theoretical investigation of the effect of plasticity on crack
  growth along a functionally graded region between dissimilar elastic-plastic
  solids}, Engineering Fracture Mechanics 69~(14-16) (2002) 1635--1645.

\bibitem{Batra2005}
R.~C. Batra, B.~M. Love, {Crack propagation due to brittle and ductile failures
  in microporous thermoelastoviscoplastic functionally graded materials},
  Engineering Fracture Mechanics 72 (2005) 1954--1979.

\bibitem{IJMMD2015}
E.~Mart{\'{i}}nez-Pa{\~{n}}eda, R.~Gallego, {Numerical analysis of quasi-static
  fracture in functionally graded materials}, International Journal of
  Mechanics and Materials in Design 11~(4) (2015) 405--424.

\bibitem{Ooi2015}
E.~Ooi, S.~Natarajan, C.~Song, F.~Tin-Loi, {Crack propagation modelling in
  functionally graded materials using scaled boundary polygons}, International
  Journal of Fracture 192~(1) (2015) 87--105.

\end{thebibliography}

\end{document}